\documentclass[aps,floatfix,twocolumn]{revtex4-2}
\usepackage[pdftex]{graphicx}
\usepackage{amsmath}
\usepackage{hyperref}
\usepackage{latexsym}
\usepackage{color}
\usepackage{amssymb}
\usepackage{amsmath}

\usepackage [latin1]{inputenc}

\usepackage{bm}
\usepackage{graphicx}
\usepackage{siunitx}
\usepackage{amsmath,amssymb}
\usepackage{braket}
\usepackage[final]{microtype}

\begin{document}

\title{Electric field dependence \\
of complex-dominated ultracold molecular collisions}

\author{Goulven Qu{\'e}m{\'e}ner}
\affiliation{Universit\'{e} Paris-Saclay, CNRS, Laboratoire Aim\'{e} Cotton, 91405 Orsay, France}
\author{James F. E. Croft}
\affiliation{The Dodd-Walls Centre for Photonic and Quantum Technologies, New Zealand}
\affiliation{Department of Physics, University of Otago, Dunedin, New Zealand}
\author{John L. Bohn}
\affiliation{JILA, NIST, and Department of Physics, University of Colorado, Boulder, Colorado 80309-0440, USA}

\begin{abstract}
Recent experiments on ultracold non-reactive dipolar molecules have observed high two-body losses, even though these molecules can undergo neither inelastic, nor reactive (as they are in their absolute ground state), nor light-assisted collisions (if they are measured in the dark).
In the presence of an electric field these losses seem to be near universal (the probability of loss at short-range is near unity) while in the absence of it the losses seem non-universal.
To explain these observations we propose a simple model based on the mixing effect of an electric field on the states of the two diatomic molecules at long-range and on the density-of-states of the tetramer complex formed at short-range, believed to be responsible for the losses.
We apply our model to collisions of ground-state molecules of endothermic systems, of current experimental interest.
\end{abstract}

\maketitle

\section{Introduction}\label{INTRO}

Recent experiments on ultracold molecules of bosonic NaRb~\cite{Guo_PRL_116_205303_2016,Ye_SA_4_eaaq0083_2018,Guo_PRX_8_041044_2018},
bosonic RbCs~\cite{Takekoshi_PRL_113_205301_2014,Molony_PRL_113_255301_2014,Gregory_NC_10_3104_2019}, 
bosonic NaK~\cite{Voges_PRL_125_083401_2020} 
and fermionic NaK~\cite{Park_PRL_114_205302_2015,Bause_PRR_3_033013_2021} 
at ultralow temperatures have observed high losses due to two-body collisions even though the molecules are
in their absolute ground state and should therefore only be able to undergo
non-lossy elastic collisions.
Several theoretical propositions have attempted to explain the origin of the losses, such as collisions with a third molecule~\cite{Mayle_PRA_87_012709_2013},
photoinduced losses~\cite{Christianen_PRL_123_123402_2019}, or
long-lived long-range roaming complexes~\cite{Klos_arXiv_2104_01625_2021}.
While in some experiments, photoinduced losses due to the trapping light
seems to explain the losses~\cite{Liu_NP_16_1132_2020,Gregory_PRL_124_163402_2020},
other experiments have questioned this possibility observing losses even where the
intensity of the trapping light is weak~\cite{Bause_PRR_3_033013_2021,Gersema_arXiv_2103_00510_2021}.
One sure thing though is that during a collision between ultracold diatomic molecules,
a long-lived tetramer complex is formed, as directly observed in a recent
experiment of reactive KRb collisions~\cite{Hu_S_366_1111_2019}.
The origin of such losses therefore remains an intriguing open question,
actively investigated from both a theoretical and experimental point of view.
The answer will certainly shed light on the role of the molecular complex
during a collision between diatomic molecules. \\

Apart from the origin of these losses another intriguing experimental feature
has been observed for non-reactive molecules:
In the presence of an electric field losses are near universal
(meaning that the probability of loss at short-range 
is nearly unity per collision)~\cite{Guo_PRX_8_041044_2018,Bause_PRR_3_033013_2021}
whereas in the absence of electric field losses are non-universal (with sub-unity probabilities per collision)~\cite{Ye_SA_4_eaaq0083_2018,Bai_PRA_100_012705_2019,
Gregory_NC_10_3104_2019,Bause_PRR_3_033013_2021}.
In other words, experiments on non-reactive molecules have shown that the
universal character for a system can depend on the applied electric field.
This is in contrast with previous experiments on reactive molecules,
especially KRb molecules~\cite{Ospelkaus_S_327_853_2010,Ni_N_464_1324_2010},
for which the system remains universal both with and without an electric
field~\cite{Quemener_PRA_81_022702_2010,Quemener_PRA_84_062703_2011,Micheli_PRL_105_073202_2010}.
This universal behavior, no matter if an electric field is applied or not,
is certainly due to the high number of channels open for reactive systems and
consequently a lower lifetime -- hence a lower importance of the complex in
this situation~\cite{Mayle_PRA_85_062712_2012,Mayle_PRA_87_012709_2013}. 

In this study, we propose a simple model that shows that even a small electric
field is sufficient to mix scattering states with different components of the
total angular momentum $J$, such that tetramer states with a high density of
states associated with higher total angular momentum start to have a significant
effect.
The model expands upon the recently developed
formalism~\cite{Croft_PRA_102_033306_2020} which compiles the concept of average cross
sections, random matrix theory, and quantum defect theory into a unified
framework to study ultracold molecular collisional processes.
The paper is organized as follows.
In Sec.~\ref{sec:concept}, we outline the concepts behind the model.
In Sec.~\ref{THEORY}, we present the model in detail and outline how the
rotational structure of the molecules in the electric field, the orbital motion,
and the use of a coupled representation can be combined to determinate the
probability $P_J$ to find a $J$ component in the total wavefunction for a
given field.
In Sec.~\ref{sec:RMT}, we extend our previous formalism to estimate
the coupling of the scattering state to the collision complex.
In Sec.~\ref{sec:FIELDEVOL}, we deduce the field evolution of the quantum defect theory
parameter describing the loss at short-range, as well as the corresponding short-range loss probability.
In Sec.~\ref{APPLICATION}, we apply this model to three systems of current
experimental interest, namely bosonic NaRb and RbCs and fermionic NaK, for which experimental data
are available.
Finally, we conclude in Sec.~\ref{CONCLU}.

\section{The concept}\label{sec:concept}

Loss of the molecules, without regard to a specific loss mechanism, can be
represented by an absorption coefficient $z_a$.
After accounting for threshold effects 
\cite{Idziaszek_PRL_104_113202_2010},
this coefficient is an energy-independent quantity that describes a
non-unitary scattering matrix ${\bar S}_{aa}$ for scattering in incident
channel $a$, via
\begin{align}
{\bar S}_{aa}^\text{abs} = \frac{ 1 - z_a }{ 1 + z_a }.
\end{align}
The absorption probability at short-range
is then  measured by the deviation of this element from
unity,
\begin{align}
{\bar p}_{\mathrm{abs}} = 1 - |{\bar S}_{aa}^{ \text{abs}}|^2 = \frac{ 4z_a }{ (1+z_a)^2 }.
\end{align}
A microscopic theory of absorption coefficients, for collisions in zero electric field,
was developed in Ref.~\cite{Croft_PRA_102_033306_2020}.
This reference distinguished absorption due to chemical reactions, 
inelastic collisions or other exothermic processes 
that are not observed in an experiment
and accounted for an overall loss. This is 
denoted by $y_a$, the unobserved absorption coeficient, associated with the short-range probability 
${\bar p}_{\mathrm{unobs}}$.
This reference also distinguished
absorption due to an indirect process, the forming of a complex (often referred to as a
``sticky collision'' due to the presence of a myriad of tetramer resonances). This is denoted by $x_a$, the indirect absorption coeficient, associated with the short-range probability 
${\bar p}_{\mathrm{res}}$.
The coefficients for these distinct processes are further combined~\cite{Croft_PRA_102_033306_2020}, their phase-shifts are added so that their scattering matrices are multiplied. As a result, one obtains
the overall absorption coefficient $z_a = (x_a+y_a)/(1 + x_a y_a)$.
For sake of simplicity, we will omit the channel index $a$ for these coefficients in the following.

\begin{figure}[t]
\begin{center}
  \includegraphics*[width=0.99\columnwidth]{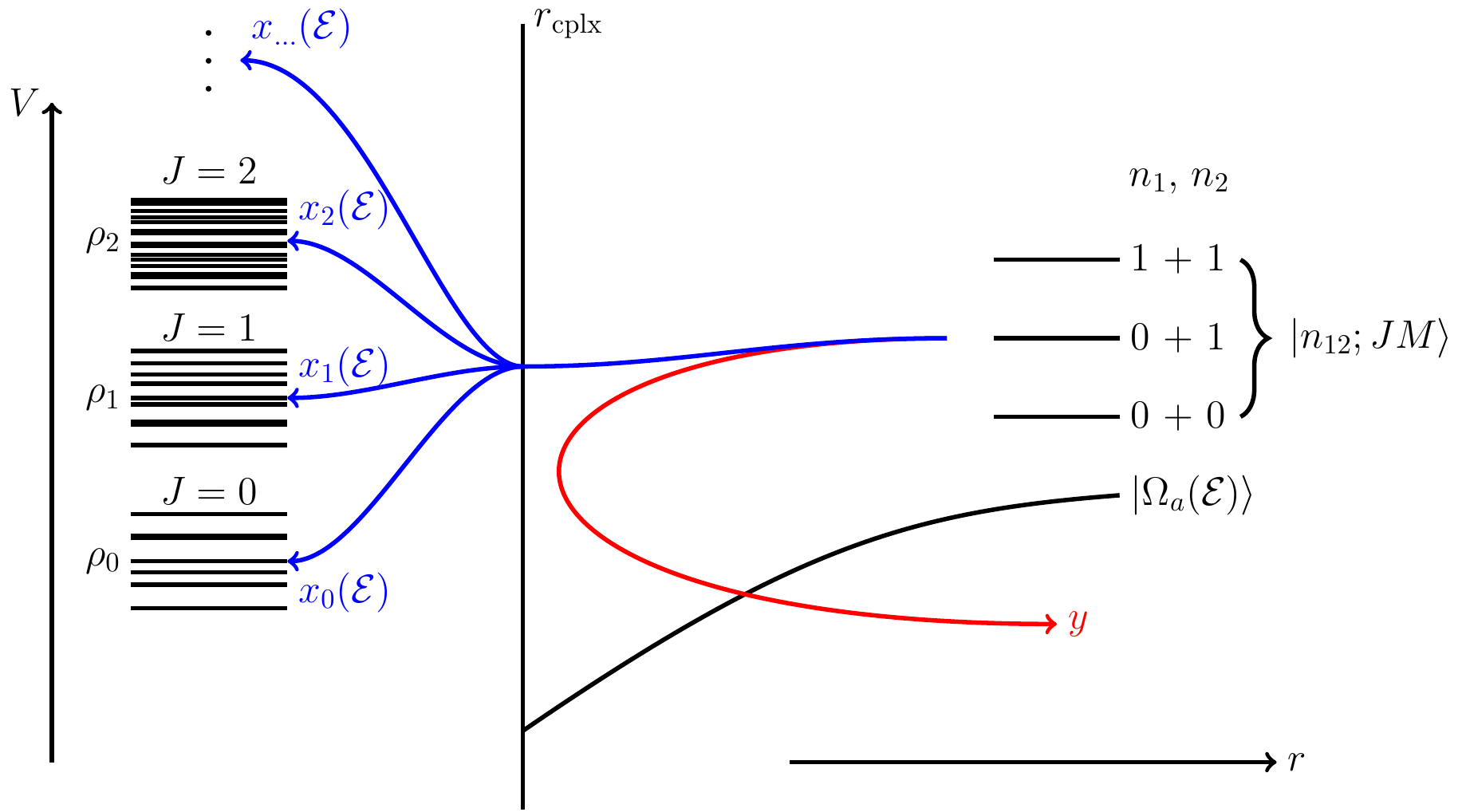}
  \caption{Schematic showing the concepts behind the model.
  At long range the field mixes the scattering states with different $n$,
  which allows for complex formation due to tetramer states at short range
  corresponding to higher $J$ than in the field free case with correspondingly higher DOS.} 
  \label{fig:schematic}
\end{center}
\end{figure}

We here adapt these ideas to nonzero electric fields ${\cal E}$, 
as depicted schematically in Fig.~\ref{fig:schematic}.
At large intermolecular distance $r$, the zero-field channels (represented by horizontal lines) are given by states with good quantum numbers of rotation $n_i$ for each molecule, and partial wave $l$.
In zero elecrtic field these are conveniently coupled to states of total angular momentum $J$ and projection $M$ as $|(n_1n_2)n_{12}l; JM \rangle$.
The value of $M$ remains a good quantum number in an electric field but all the others, in particular $J$, are good only in zero field.

When the field is turned on,
we assume that the $J$-mixing at short-range
is solely due to the long-range physics, that is 
the electric field seen by the individual molecules and their dipole-dipole interaction.
We consider that the electric field is small enough so that 
the short-range physics and the potential energy surface are not affected
and remain unchanged through the process.
The lowest adiabatic channel becomes a superposition of the zero-field channels, hence it consists of states of different values of angular momentum $J$.
This adiabatic channel is denoted $|\Omega_a({\cal E}) \rangle$ in the figure.
The coupling of this initial scattering state to the different $J$ states of the complex is assumed to occur at the characteristic scale of the complex, $r_\mathrm{cplx}$.
Following our assumption and as we will see later using a two-level model, 
the electric field at which multiple angular momentum states become relevant is 
typically governed by the field at which the dipole-dipole interaction is comparable to the centrifugal energy at this radius, namely
\begin{eqnarray}
\frac{d_\mathrm{ind}({\cal E})^2}{4 \pi \epsilon_0 r_\mathrm{cplx}^3} \sim \frac{\hbar^2}{2 \mu r_\mathrm{cplx}^2}.
\end{eqnarray}
Here $\mu$ is the reduced mass of the scattering partners, 
and the induced dipole moment $d_\mathrm{ind}$ is 
the expectation value of the permanent dipole moment $d$ at this electric field.
For the molecules relevant to experiments, as we will see, this field is on the order of 100~V/cm.

Upon colliding the molecules can meet with various fates which we regard as independent, in the same sense as in~\cite{Croft_PRA_102_033306_2020}.
In the figure this includes the possibility for unobserved scattering processes with the coefficient $y$.
It also includes several processes where the molecules vanish due to the formation of complexes with well-defined values of $J$ and density of states $\rho_J$, coupled to the entrance channel by the corresponding 
indirect absorption coefficients $x_J({\cal E})$.
A critical assumption is that the resonant tetramer states with different $J$ values are not coupled by the electric field, due to physical arguments given below.
Therefore, in zero field the molecules enter states of the complex with a well-defined total angular momentum $J$, but in a field they can access additional states, leading to increased opportunities for 
complex formation via sticking.

Realizing the model therefore consists of three parts.
First we assess the influence of the electric field in mixing the zero-field 
states into the adiabatic channel $|\Omega_a({\cal E}) \rangle$, next we evaluate the indirect absorption coefficients $x_J({\cal E})$ for this state to the various angular momentum states of the complex,
and finally we combine these together to obtain the final field-dependent absorption coefficient.
As in this study we will consider the case of ultracold, endothermic ground states molecules, 
no chemical reactions nor inelastic collisions will occur and there will be no coefficient $y$.
Therefore, the $z$ coefficient reduces to the $x$ coefficient and the 
short-range absorption probability ${\bar p}_{\mathrm{abs}}$ responsible for the overall loss
identifies with ${\bar p}_{\mathrm{res}}$.

\section{Channel mixing due to the electric field}

\label{THEORY}

We consider  identical diatomic molecules characterized by their rotational states $|n \ m_n\rangle$, with $n=0$, 1,\dots.
Electronic and nuclear spins are considered as spectators, fixed to identical values in each molecule and unchanged by the collision.
At ultralow temperatures, identical molecules 
in indistinguishable states collide in their lowest 
available partial wave: $l=0$ for bosons and $l=1$ for fermions.
In the presence of an electric field $\vec{{\cal E}}$, states with different $n$ are mixed, giving them laboratory-frame, 
induced dipole moments.
These moments in turn couple different values of $l$.
Scattering is therefore best described using field dressed states, which we construct in this section.
The first part describes the rotational structure of the individual molecules in an electric field;
the second part describes the orbital angular momentum coupling;
the third part uses the transformation from an uncoupled to a coupled representation to determine the probability to find a $J$ component in the incident channel wavefunction for a given electric field.

\subsection{Field-dressed states of the molecules}

The transition from non-universal to universal loss occurs at electric fields that are perturbative with respect to mixing the rotational states of the molecules.
As we are interested in collisions of molecules in their ground states, we therefore approximate the field dressed state as the $n=0$ ground state, perturbed by the first rotational excitation $n=1$.
These states have the unperturbed energies $0$ and $2B$, respectively, where $B$ is the rotational constant.
The electric field Hamiltonian $H_S = -\vec{d}.\vec{{\cal E}}$ mixes these states.
The expression of the Stark effect in the rotational basis
is given by~\cite{Wang_NJP_17_035015_2015}
\begin{multline}
 \langle {n} \ m_{n} | H_S| {n}' \ m_{n}' \rangle  =
  - d \,  {\cal E} \, \delta_{m_n,m_n'} \, (-1)^{m_{n}} \times \\
  \sqrt{(2n+1) \, (2n'+1)} \,
  \left(\begin{array}{ccc} n & 1 & n' \\ 0 & 0 & 0 \end{array} \right)
\, \left(\begin{array}{ccc} n & 1 & n' \\ -m_{n} & 0 & m_{n}' \end{array}  \right)
\end{multline}
giving the Hamiltonian in matrix form, for $m_n=0$,
\begin{eqnarray}
\left[ \begin{array}{cc}
 0 &
-d \,  {\cal E} /\sqrt{3}   \\
-d\,  {\cal E} /\sqrt{3} &
2B
\end{array}\right] .
\end{eqnarray}
The lowest eigenstate of this Hamiltonian is the molecular ground state of interest, denoted by
\begin{eqnarray}
\big| \tilde{0} \big\rangle &=& \alpha \, \big| {0} \big\rangle + \beta \, \big| {1} \big\rangle
\label{ES1mol}
\end{eqnarray}
with
\begin{gather}
\alpha = \cos(\theta/{ 2}) \qquad \beta = -\sin(\theta/{ 2})  \nonumber \\
\theta = \arctan (d \,  {\cal E}  / { \sqrt{3}}B) .
\end{gather}
In the following, $\big| \tilde{0} \big\rangle $ is referred
to as the dressed state (that is dressed by the electric field),
as opposed to the bare state $\big| 0 \big\rangle $ in zero electric field.
The corresponding eigenenergy is
\begin{equation}
E_{\tilde{0}} = B - \sqrt{B^2 + (d \,  {\cal E} / {\sqrt{3}})^2}
\label{EE1mol}
\end{equation}
These expressions are valid as long as
the next rotational state $n=2$ of rotational energy $6B$ remains only weakly coupled to the $n=1$ state of energy $2B$. This is the case when $|\big\langle 1 \, 0 \big|H_S \big| 2 \, 0 \big\rangle| \ll 6B - 2B$, that is when
$d \, {\cal E}/B \ll 2 \sqrt{15} \simeq 7.75$.
As we will see, this limit is satisfied for the range for fields over which the
transition from non-universal to universal behavior occurs.

\subsection{Collisional channel in an electric field}

Generally, the dipole-dipole interaction is usefully computed in an uncoupled basis set
\begin{eqnarray}
\big| n_1 \ m_{n_1} \big\rangle \, \big| n_2 \ m_{n_2} \big\rangle \, \big| l \ m_l \big\rangle,
\label{CMS-UB}
\end{eqnarray}
where $l$ is the orbital angular momentum and $m_l$ is its laboratory-frame projection.
This basis is denoted the combined molecular state (CMS).
It defines the bare channels, in zero field, and its quantum numbers are good in zero field and in the limit where the molecules are far apart.
Because we intend to connect the scattering states to states of the complex that have particular values of total angular momentum $J$, it will be useful at the last stage to recombine the CMS basis into a total angular momentum representation,
\begin{align}
|(n_1 n_2)n_{12}l;JM \rangle \equiv | \lambda; JM \rangle
\end{align}
by the usual rules of angular momentum coupling.  Here $\lambda$ is introduced as a shorthand notation and a reminder of the coupling scheme.

If the molecules are dipolar, matrix elements of the dipole-dipole interaction 
in the uncoupled basis
are given by the general expression~\cite{Wang_NJP_17_035015_2015}
\begin{multline}
 \langle {n}_1 \ m_{n_1}, {n}_2 \ m_{n_2}, l \ m_l  | V_\mathrm{dd} | {n}_1' \ m_{n_1}', {n}_2' \ m_{n_2}', l' \ m_l' \rangle  = \\
  - \sqrt{30} \, \frac{d^2}{4 \pi \varepsilon_0 r^3} \times \\
  \sum_{m_{\lambda_1}=-1}^{1}  \sum_{m_{\lambda_2}=-1}^{1}   \sum_{m_\lambda=-2}^{2}
 (-1)^{m_{n_1}+m_{n_2}+m_l}
 \, \left( \begin{array}{ccc}  1 & 1 & 2 \\ 0 & 0 & 0  \end{array}  \right) \\
\times  \sqrt{(2n_1+1) \, (2n_1'+1)}
\, \left(\begin{array}{ccc} n_1 & 1 & n_1' \\ 0 & 0 & 0 \end{array} \right)
\, \left(\begin{array}{ccc} n_1 & 1 & n_1' \\ -m_{n_1} & m_{\lambda_1} & m_{n_1}' \end{array}  \right)  \\
\times \sqrt{(2n_2+1) \, (2n_2'+1)}
\, \left(\begin{array}{ccc} n_2 & 1 & n_2' \\ 0 & 0 & 0 \end{array} \right)
\, \left(\begin{array}{ccc} n_2 & 1 & n_2' \\ -m_{n_2} & m_{\lambda_2} & m_{n_2}' \end{array}  \right) \\
 \times  \sqrt{(2l+1) \, (2l'+1)}
\, \left(\begin{array}{ccc} l & 2 & l' \\ 0 & 0 & 0 \end{array} \right)
\, \left(\begin{array}{ccc} l & 2 & l' \\ -m_{l} & -m_{\lambda} & m_{l}' \end{array}  \right).
\label{Vdd}
\end{multline}
For collisions of molecules in the dressed ground state $|{\tilde 0} \rangle$, we require matrix elements of the dipole-dipole interaction in the basis $|{\tilde 0}, {\tilde 0}, l\rangle$.
These are easily computed from the matrix elements of this interaction in the lowest dressed state of the molecule as described by Eq.~\eqref{ES1mol}.
Note that, since the nuclear spin is regarded as a spectator degree of freedom and identical in both molecules, these states are already symmetric under particle exchange for even $l$, and antisymmetric for odd $l$.

Finally, within the model we restrict attention to just the two lowest relevant partial waves, and to components $m_l=0$ of these partial waves.
This will afford an analytical representation of the adiabatic channel and its relation to the total angular momentum representation.
This calculation is carried out in Appendix~\ref{app:a} and summarized in the following subsections.

\subsubsection{Bosons}

For identical bosons in indistinguishable states, the two relevant channels are $\big| \tilde{0} ,  \tilde{0} , 0\rangle$
and $\big| \tilde{0} ,  \tilde{0} , 2\rangle$.
In the absence of the dipole-dipole interaction these have energies
$2 E_{\tilde{0}} - C^\mathrm{el}_6/r^6$ and
$2 E_{\tilde{0}} - C^\mathrm{el}_6/r^6 + 6 \hbar^2 / 2 \mu r^2$, respectively, where
$C^\mathrm{el}_6$ is the
coefficient of the electronic contribution
of the van der Waals interaction~\cite{Lepers_PRA_88_032709_2013}
($C^\mathrm{el}_6$ is a positive number for molecules in their ground state).
Note that the rotational contribution of the van der Waals coefficient is automatically included in our model, as it arises from the perturbations due to the dipole-dipole interaction.  %
The coupling between these two channels in the dressed basis is given by (see Appendix~\ref{app:a})
\begin{align}
 \langle \tilde{0} , \tilde{0}, 0 | V_\mathrm{dd} | \tilde{0} , \tilde{0}, 0 \rangle
 =  0,
\end{align}
\begin{align}
 \langle \tilde{0} , \tilde{0}, 2 | V_\mathrm{dd} | \tilde{0} , \tilde{0}, 2 \rangle
=
- \frac{d^2}{4 \pi \varepsilon_0 r^3}
 \,  4 \, \alpha^2 \beta^2 \, (4/21),
\end{align}
and
\begin{align}
  \langle \tilde{0} , \tilde{0}, 2 | V_\mathrm{dd} | \tilde{0} , \tilde{0}, 0 \rangle &=
  \langle \tilde{0} , \tilde{0}, 0 | V_\mathrm{dd} | \tilde{0} , \tilde{0}, 2 \rangle
 \nonumber \\
  &= - \frac{d^2}{4 \pi \varepsilon_0 r^3}
  \,  4 \, \alpha^2 \beta^2 \, (2  / \, 3\sqrt{5}) .
\end{align}
The lowest eigenvalue of this two-by-two matrix defines the adiabatic channel of interest, denoted
\begin{eqnarray}
\big| \Omega_a({\cal E}) \big\rangle \equiv \big| \tilde{0} ,  \tilde{0}, \tilde{0} \big\rangle .
\label{LOWESTSTATE-B}
\end{eqnarray}
The third symbol ${\tilde 0}$ is a reminder that the partial wave $l=0$ is no longer strictly good, but is dressed by the dipole interaction.
This represents an adiabatic state whose value varies with $r$, as follows.
Define ${E}_1 =  2 E_{\tilde{0}} - {C_6^\mathrm{el}} / {r^6} $ and
${E}_2 = 2 E_{\tilde{0}} - {C_6^\mathrm{el}} / {r^6} + {6 \hbar^2} / {2  \mu  r^2} - ({d^2} / {4 \pi \varepsilon_0 r^3}) \,  4 \, \alpha^2 \beta^2 \, (4/21) $.
For a given ${\cal E}$, if ${E}_2 \ge {E}_1$, this channel is given by
\begin{eqnarray}
\big| \Omega_a({\cal E}) \big\rangle =
\cos(\eta{/2})  \, \big| \tilde{0} ,  \tilde{0}, {0} \big\rangle
-\sin(\eta{/2})  \, \big| \tilde{0} ,  \tilde{0}, 2 \big\rangle
\nonumber \\
\end{eqnarray}
and if ${E}_2 < {E}_1$, it is given by
\begin{eqnarray}
\big| \Omega_a({\cal E}) \big\rangle =
-\sin(\eta{/2}) \, \big| \tilde{0} ,  \tilde{0}, {0} \big\rangle
+ \cos(\eta{/2})   \, \big| \tilde{0} ,  \tilde{0}, 2 \big\rangle,
\nonumber \\
\end{eqnarray}
with mixing angle
\begin{eqnarray}
\eta = \arctan \bigg\{ \frac{ { \sin^2 \theta} \, (4  / \, 3\sqrt{5})}{ \big| 6 \, \tilde{r}
- { \sin^2 \theta} \, (4/21) \big| } \bigg\} .
\label{eta-B}
\end{eqnarray}
We used the fact that $ 4 \, \alpha^2 \beta^2 = { \sin^2 \theta}$.
The rescaled length
\begin{eqnarray}
\tilde{r} = \frac{r}{a_\mathrm{dd}}
\end{eqnarray}
has been introduced
with
\begin{eqnarray}
 a_\mathrm{dd} = \frac{2 \mu}{\hbar^2} \, \bigg(\frac{d^2}{4 \pi \epsilon_0}\bigg)
\end{eqnarray}
being the characteristic dipole-dipole length~\cite{Gao_PRA_78_012702_2008}.
For the case of non-reactive bi-alkali dipolar molecules, $a_\mathrm{dd} \sim [10^5-10^6] \ a_0$ 
\cite{Gonzalez-Martinez_PRA_96_032718_2017}. The couplings are estimated at 
$r = r_\text{cplx}$. A typical value for the length scale of the complex is 
$r_\text{cplx} \sim 5 \ \text{{\r{A}}} \sim 10 \ a_0$ \cite{Byrd_PRA_82_010502_2010,Yang_JPCL_11_2605_2020,Christianen_JCP_150_064106_2019,Klos_arXiv_2104_01625_2021}. 
Then, $\tilde{r} \sim [10^{-5} - 10^{-4}]$. 
$\big| \Omega_a({\cal E}) \big\rangle $ in
Eq.~\eqref{LOWESTSTATE-B} is the expression of the lowest dressed channel of the system for bosonic molecules within the model.

\subsubsection{Fermions}

We proceed similarly for identical fermions.
The two relevant channels are now
$\big| \tilde{0} ,  \tilde{0} , 1\rangle$
and $\big| \tilde{0} ,  \tilde{0} , 3\rangle$,
whose energies exclusive of the dipole-dipole interaction are
$2 E_{\tilde{0}} - C^\mathrm{el}_6/r^6 + 2 \hbar^2/2 \mu r^2$ and
$2 E_{\tilde{0}} - C^\mathrm{el}_6/r^6 + 12 \hbar^2/2 \mu r^2$.
The coupling matrix elements are (see Appendix~\ref{app:a})
\begin{eqnarray}
  \langle \tilde{0} , \tilde{0}, 1 | V_\mathrm{dd} | \tilde{0} , \tilde{0}, 1 \rangle
= - \frac{d^2}{4 \pi \varepsilon_0 r^3} \, 4 \, \alpha^2 \beta^2 \, (4/15),
\end{eqnarray}
\begin{eqnarray}
\langle \tilde{0} , \tilde{0}, 3 | V_\mathrm{dd} | \tilde{0} , \tilde{0}, 3 \rangle
= - \frac{d^2}{4 \pi \varepsilon_0 r^3}
 \,  4 \, \alpha^2 \beta^2 \, (8/45), \nonumber \\
\end{eqnarray}
and
\begin{align}
 \langle \tilde{0} , \tilde{0}, 3 | V_\mathrm{dd} | \tilde{0} , \tilde{0}, 1 \rangle &=
  \langle \tilde{0} , \tilde{0}, 1 | V_\mathrm{dd} | \tilde{0} , \tilde{0}, 3 \rangle \nonumber \\
&= - \frac{d^2}{4 \pi \varepsilon_0 r^3}
 \,   4 \, \alpha^2 \beta^2 \, (2 \sqrt{3} \, / \, 5\sqrt{7}) . \nonumber \\
\end{align}
The new eigenstate of the lowest channel is then represented by
\begin{eqnarray}
\big| \Omega_a({\cal E}) \big\rangle \equiv \big| \tilde{0} ,  \tilde{0}, \tilde{1} \big\rangle .
\label{LOWESTSTATE-F}
\end{eqnarray}
Proceeding as before, define
${E}_1 = 2 E_{\tilde{0}} - {C_6^\mathrm{el}}/{r^6} + {2 \hbar^2}/{2  \mu  r^2} - ({d^2}/{4 \pi \varepsilon_0 r^3})  \, 4 \, \alpha^2 \beta^2 \, (4/15) $ and
${E}_2 = 2 E_{\tilde{0}} - {C_6^\mathrm{el}}/{r^6} + {12 \hbar^2}/{2  \mu  r^2} - ({d^2}/{4 \pi \varepsilon_0 r^3}) \, 4 \, \alpha^2 \beta^2 \, (8/45) $.
For a given ${\cal E}$, if ${E}_2 \ge {E}_1$, the channel is given by
\begin{eqnarray}
\big| \Omega_a({\cal E}) \big\rangle =
\cos(\eta{/2})  \, \big| \tilde{0} ,  \tilde{0}, {1} \big\rangle
-\sin(\eta{/2})  \, \big| \tilde{0} ,  \tilde{0}, 3 \big\rangle
\nonumber \\
\end{eqnarray}
and if ${E}_2 < {E}_1$, it is given by
\begin{eqnarray}
\big| \Omega_a({\cal E}) \big\rangle =
-\sin(\eta{/2}) \, \big| \tilde{0} ,  \tilde{0}, {1} \big\rangle
+ \cos(\eta{/2})   \, \big| \tilde{0} ,  \tilde{0}, 3 \big\rangle ,
\nonumber \\
\end{eqnarray}
with mixing angle
\begin{eqnarray}
\eta = \arctan \bigg\{ \frac{ { \sin^2 \theta} \, (4 \sqrt{3} \, / \, 5\sqrt{7})}{\big| 10 \, \tilde{r}
+ { \sin^2 \theta} \, (4/45) \big| } \bigg\}  .
\label{eta-F}
\end{eqnarray}
Similarly, $\big| \Omega_a({\cal E}) \big\rangle$ in Eq.~\eqref{LOWESTSTATE-F} is the expression of the lowest dressed channel of the system
for fermionic molecules within the model.

\subsection{Probability to find a $J$ component}

From these expressions for the lowest dressed channels in the uncoupled representation,
one can transform them into the coupled, total angular momentum representation.
It is then straightforward to extract the probability $P_J$ to find a
$\big| J \, M \big\rangle  = \big| J \ 0 \big\rangle$ component contained in the
wavefunction $\big| \Omega_a({\cal E}) \big\rangle$ of the dressed channel when an electric field is turned on.
This probability is evaluated at the specific value of $\tilde{r} = r_{\text{cplx}} / a_\mathrm{dd}$,
that is at a position around the characteristic scale of the tetramer complex.
This probability is given by
\begin{align}
P_{J}({\cal E}) &=  \sum_{(n_1, n_2) \, n_{12}, \, l } \big| \big\langle  (n_1, n_2) \, n_{12}  \, l \, ; J \ 0 \big| \Omega_a({\cal E}) \big\rangle \big|^2 \nonumber \\
&= \sum_{\lambda} |\langle \lambda; J0 | \Omega_a({\cal E}) \rangle |^2
\label{PJ}
\end{align}
where the sum runs over all combinations of $(n_1, n_2) \, n_{12}, \, l $ consistent
with the given $J$ under angular momenta couplings.
The expressions of the probabilities $P_{J}^{\text{B,F}}({\cal E})$ for each $J$ for bosons or fermions
are provided in Appendix~\ref{app:c}.
Note that within the restrictions of the model,  $n=0,1$ and $l=0,2$ implies $J\le4$ for bosons; while $l=1,3$ implies  $J\le 5$ for fermions.

\begin{figure}[t]
\begin{center}
\includegraphics*[width=8cm, trim=0cm 0cm 0cm 0cm]{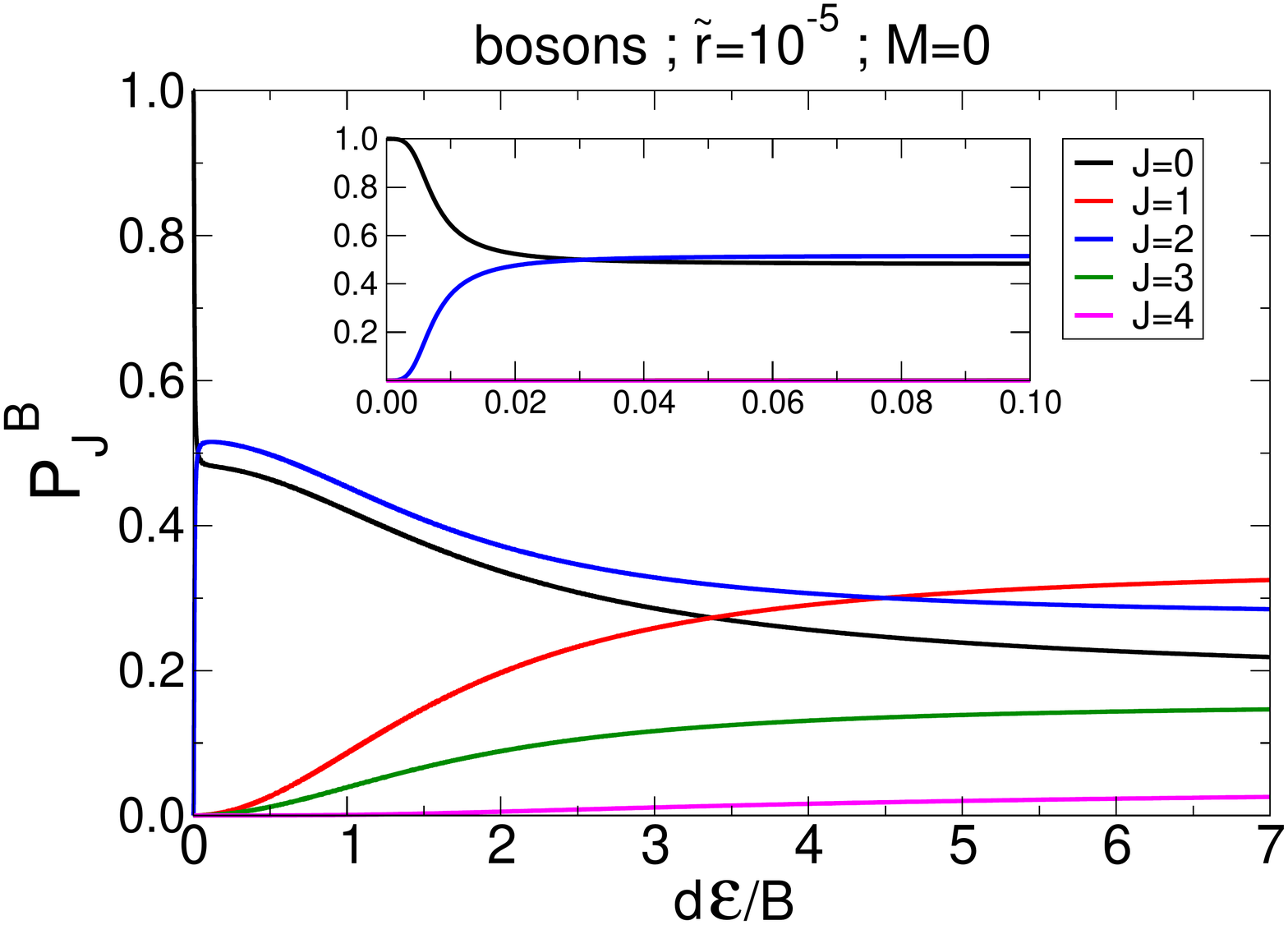} \\
\includegraphics*[width=8cm, trim=0cm 0cm 0cm 0cm]{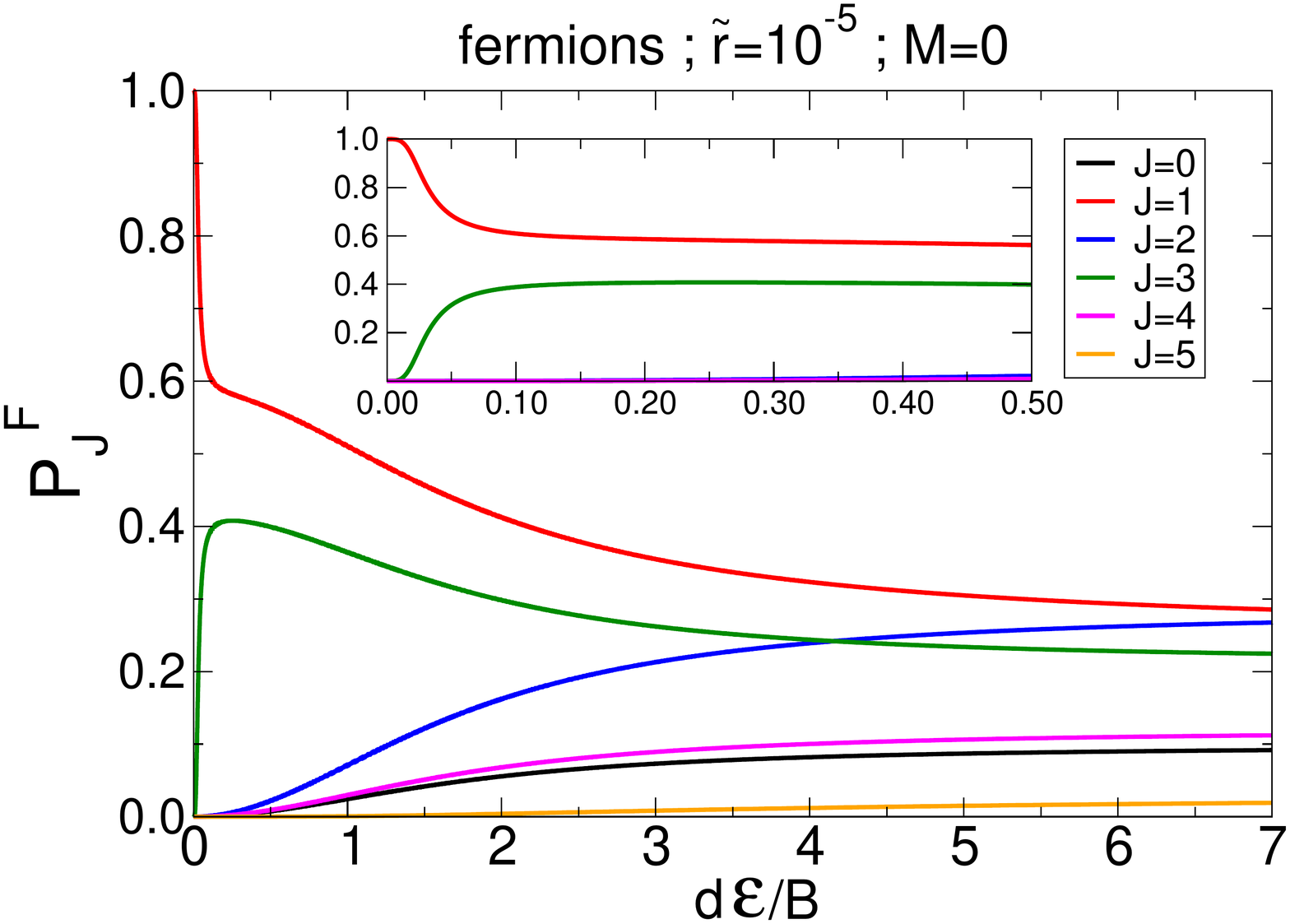}
\caption{Probabilities $P_J^\text{B}$ for bosons (top panel) and $P_J^\text{F}$ for fermions (bottom panel)
versus $d  {\cal E} / B$ at a fixed value of $\tilde{r}=10^{-5}$. The contributions of $J=0,1,2,3,4,5$ are plotted respectively in black, red, blue, green, pink, orange.
The model considers only the projection $M=0$.
The insets provide a close-up of the figure at low values of $d  {\cal E} / B$.
Typically, one unit of $d  {\cal E} / B$ corresponds to an electric field of ${\cal E} \sim $~0.8~kV/cm, ${\cal E} \sim $~1.2~kV/cm and ${\cal E} \sim $~1.97~kV/cm for RbCs, NaRb and NaK respectivelly 
\cite{Gonzalez-Martinez_PRA_96_032718_2017}.
}
\label{FIG-PROBA-ADIM}
\end{center}
\end{figure}

We plot these probablities in Fig.~\ref{FIG-PROBA-ADIM}
for bosons (top panel) and for fermions (bottom panel) for a fixed value of
$\tilde{r}=10^{-5}$, which is a typical value for non-reactive bi-alkali dipolar molecules
(see above).
As can be seen in the inset of the figures and from the equations in Appendix~\ref{app:c},
the $J=0$ for bosons ($J=1$ for fermions) is the main and only contribution when ${\cal E}=0$,
as expected for a $l=0$ s-wave ($l=1$ p-wave) collision in the absence of an electric field.
But once ${\cal E}$ is turned on, the $J=2$ ($J=3$) component
increases and becomes of similar magnitude with the $J=0$ ($J=1$) contribution.
Therefore, a small applied electric field is sufficient to significantly couple
the rotational $n_1, n_2$ and orbital $l$ angular momenta of the system,
creating other possible contributions of $J$ to the one dominating at zero electric field.
As the density-of-states of tetramer bound states at short range depends
on the value of the total angular momentum $J$~\cite{Christianen_PRA_100_032708_2019},
this figure shows qualitatively why such a small field is required for the
non-universal to universal loss transition to occur.

Note that there is a difference between the two types of species. 
While for the bosonic case
the probability curves cross for $J=0$ and $J=2$,
the corresponding crossing of $J=1$ and $J=3$ does not occur for fermions.
This is due to the fact that for bosons,
${E}_2 - {E}_1 = (d^2 / \, 4 \pi \varepsilon_0 r^3) \, (6 \, r/a_\mathrm{dd} - { \sin^2 \theta} \, (4/21))$ can become negative for a particular electric field ${\cal E}={\cal E}^*$ and/or a position $r^*$
and then the lowest dressed eigenstate changes of main character,
from $\big| \tilde{0} ,  \tilde{0}, {0} \big\rangle$
to $\big| \tilde{0} ,  \tilde{0}, {2} \big\rangle$ in Eq.~\eqref{LOWESTSTATE-B}.
If we fix $r \simeq r_\text{cplx}$,
$P_2^B$ becomes greater than $P_0^B$ at a given electric field, as seen in Fig~\ref{FIG-PROBA-ADIM}.
The corresponding value of $d \, {\cal E}^* / B$ is the one that satisfies
$6 \, r/a_\mathrm{dd} = 6 \, \tilde{r} = {\sin^2 \theta} \, (4/21)$
that is $d \, {\cal E}^* / B = \sqrt{3} \, \tan(\arcsin{\sqrt{63 \, \tilde{r} / 2}})$.
Here for $\tilde{r} = 10^{-5}$, $d \, {\cal E}^* / B  \simeq 0.03$ which is in agreement with the crossing seen in the inset.
This is not the case for fermions as ${E}_2 - {E}_1 = (d^2 / \, 4 \pi \varepsilon_0 r^3) \, (10 \, r/a_\mathrm{dd} + {\sin^2 \theta} \, (4/45))$ is always positive
and the main character of the lowest dressed eigenstate remains $\big| \tilde{0} ,  \tilde{0}, {1} \big\rangle$
all along, so that $P_1^F$ remains greater than $P_3^F$.

\section{Coupling to the collision complex}

\label{sec:RMT}

The probability $P_J({\cal E})$ for the incident molecules to find a collision complex of angular  momentum $J$ plays a critical role in the short-range probability $\bar{p}_{\text{res}}$, as we now explore.  We begin by
reviewing the zero-field case of our recent unified model~\cite{Croft_PRA_102_033306_2020}. In zero electric field, asymptotic states of the scattering wave function consist of the terms
\begin{align}
|\Psi_{\lambda}^{JM} \rangle = r^{-1} \, \psi_{\lambda}^{JM}(r) \, | \lambda; JM \rangle,
\end{align}
where $\psi_{\lambda}^{JM}$ is the solution to the radial Schr\"odinger equation in channel $|\lambda; JM \rangle$, at the total energy $E$.
Each bound state of the complex is a highly multichannel wave function, denoted $\mu$  (not to mistake with the reduced mass here) but expanded into a convenient channel basis $|i \rangle$, as
\begin{align}
| \Phi_{\mu}^{JM} \rangle = r^{-1} \, \sum_i  \phi_i^{JM}(r) \, | i; JM \rangle.
\end{align}
This state of course also preserves $J$ in zero electric field.
Coupling between the continuum states and the states of the complex is mediated by matrix elements of a potential energy $V(r)$, as
\begin{align}
\langle \Psi_{\lambda}^{JM} | V(r) | \Phi_{\mu}^{J^{\prime}M^{\prime}} \rangle \equiv W_{\lambda\mu}^{JM}
\delta_{JJ^{\prime}} \delta_{MM^{\prime}} .
\end{align}
Writing this more completely we have
\begin{equation}
W_{\lambda \mu}^{JM} = \int dr \, \psi_{\lambda}^{JM*}(r) \, \sum_i  \phi_i^{JM} (r) \, \langle \lambda; JM | V(r) | i ; JM \rangle.
\label{eq:W_zero_field}
\end{equation}

In the statistical theory of resonant states these matrix elements are random variables sampled from the Gaussian distribution characterized by a variance, which becomes a parameter of the theory.
For chaotic scattering of molecules it is known that, separately, the radial coupling functions $\langle \lambda; JM | V(r) | i ; JM \rangle$ are Gaussian distributed \cite{Croft_NC_8_15897_2017}.
In addition to this, the multiple radial integrals in Eq.~\eqref{eq:W_zero_field} consist of integrals over oscillating functions and can be considered to merely contribute to the overall distribution.
As such we assume that there is no correlation between the various states $\lambda$ and $\mu$.  
Within the statistical model we therefore follow \cite{Mitchell_RMP_82_2845_2010}
and treat the matrix elements as Gaussian distributed random variables, characterized by a mean coupling constant
\begin{align}
\langle W_{\lambda \mu}^{JM} W_{\nu \lambda^{\prime}}^{JM} \rangle
&= \delta_{\mu \nu} \delta_{\lambda \lambda^{\prime}} \left( \nu_{\lambda;JM} \right)^2  \nonumber \\
& \equiv \delta_{\mu \nu} \delta_{\lambda \lambda^{\prime}} \nu_J^2. \label{eq:meancc}
\end{align}
The last line incorporates another approximation, that the variance is independent of the bare open channel.
The delta functions mean that the coupling between states with different quantum numbers are uncorrelated.
These $\nu_{J}$'s determine the indirect absorption coefficient
\begin{align}
x_{J} = \frac{ \pi^2 \nu_{J}^2 }{ d_J } ,
\end{align}
where $d_J$ is the mean spacing between levels of the complex with angular momentum $J$,
corresponding to a density-of-states $\rho_J = 1/d_J$.
Within the statistical theory, $x_{J}$ is the coefficient used to assess 
the short-range probability $\bar{p}_{\text{res}}$
\cite{Croft_PRA_102_033306_2020}.

When the electric field is nonzero,  we must asses the influence of the field on both the states of the incident channel and the complex.
Here we make a key assumption, that the states of the complex $|\Phi_{\mu}^{JM} \rangle$ remain states of good angular momentum, and  that, although their energy levels and matrix elements can change, these changes do not affect the overall mean spacing $d_J$ or the strength of the coupling $\nu_J^2$.
This assumption is justified by detailed studies of the potential energy surface for the reactive KRb system~\cite{Byrd_PRA_82_010502_2010,Yang_JPCL_11_2605_2020} as well as the non-reactive NaK~\cite{Christianen_JCP_150_064106_2019}
and NaRb~\cite{Klos_arXiv_2104_01625_2021} systems.
The trends seen there can be extended to all other combinations of bi-alkali molecules~\cite{Byrd_JCP_136_014306_2012}.
Specifically, the collisional entrance channel between two molecules
of type AB + AB corelates to two geometries of D$_{2h}$ symmetries
(D$_{2h}$-I, D$_{2h}$-II) of the A$_2$B$_2$ tetratomic bound state. These symmetries are quite special as they equally
put two similar atoms from each side of the  $x$ and $y$ axis in the plane where the tetramer stands,
resulting in an equal but opposed electronic charge distribution of those atoms,
and an automatic cancelation of the overall dipole moment of the tetramer in its body-fixed frame.
Note that for NaK~\cite{Christianen_JCP_150_064106_2019}, a C$_{2h}$ symmetry was found
for the second minimum instead of the D$_{2h}$-II. 
But similar arguments still hold.
Therefore, given that the entrance channel is AB + AB, we assume the tetramer bound states
 are not mixed by the field because there is no permanent dipole moment
for this geometry.
On the other hand, having crossed the transition state, the tetramer can find itself in a state of C$_s$ symmetry, where this cancellation of dipoles does not hold.
We therefore assume that the act of complex forming represented by the coefficients $x_J$ implicitly represents the \emph{initial} stages of complex forming in the D$_{2h}$ region of the potential energy surface, with the C$_s$ region relevant to the further time evolution of the complex.
The limitations of these assumptions should of course be tested in further elaborations 
of the present theory.

The entire meaningful influence of the electric field is therefore assumed to be its influence on the incident channel $|\Psi_a({\cal E}) \rangle$.  This state is given by
\begin{align}
|\Psi_a({\cal E}) \rangle = r^{-1} \,  \omega_a({\cal E};r) \, |\Omega_a({\cal E})  \rangle
\end{align}
where $|\Omega_a ({\cal E}) \rangle$ is the adiabatic wave function defined in the previous section, and $\omega_a(r)$ is the radial function in the corresponding adiabatic potential.
The absorption probability from the adiabatic channel to the states of the complex are again governed by the coupling matrix elements
\begin{align}
W_{a \mu}^{JM}({\cal E}) &= \langle \Psi_a({\cal E}) | V(r) | \Phi_{\mu}^{JM} \rangle \\
&=  \int dr \, \omega_a({\cal E};r) \, \sum_i \phi_i^{JM}(r) \,
\langle \Omega_a({\cal E}) | V(r) | i; JM \rangle \nonumber
\end{align}
where we have used the assumption that the states of the complex are the same as in zero field.
Because the adiabatic function $\omega_a$ is of the same magnitude as the diabatic radial functions $\psi_{\lambda}$ in Eq.~\eqref{eq:W_zero_field}, 
the influence of the radial functions on the statistics of the matrix elements are the same for the diabatic function as for the adiabatic functions.

The change in the statistics of the  matrix elements $W$ resides therefore entirely  in the  channel coupling  matrices $\langle \Psi_a({\cal E}) |  V(r) | \Phi_{\mu}^{JM} \rangle$.  Setting $M=0$ for the model at hand, we have
\begin{align}
W_{a \mu}^{J0}({\cal E}) &= \sum_{\lambda,J^{\prime},M^{\prime}}
\langle \Psi_a({\cal E})  | \Psi_{\lambda}^{J^{\prime}M^{\prime}} \rangle \,
\langle \Psi_{\lambda}^{J^{\prime}M^{\prime}} | V(r) | \Phi_{\mu}^{JM} \rangle \nonumber \\
&= \sum_\lambda \langle \Omega_a({\cal E}) | \lambda; J0 \rangle \, W_{\lambda \mu}^{J0}.
\end{align}
Using the statistical properties of the zero-field matrix elements $W_{\lambda \mu}^{JM}$ 
from Eq.~\eqref{eq:meancc}, we find the variance of the  matrix elements at non-zero field
\begin{align}
  \langle W_{a \mu}^{J0}({\cal E}) & \, W_{\nu a}^{J0}({\cal E}) \rangle \nonumber \\
& = \sum_{\lambda \lambda^{\prime}} \langle \Omega_a({\cal E}) | \lambda;J0 \rangle
\, \langle W_{\lambda \mu}^{J0} \, W_{ \nu \lambda^{\prime}}^{J0} \rangle
\, \langle \lambda^{\prime}; J0 | \Omega_a({\cal E}) \rangle \nonumber \\
&= \sum_{\lambda} | \langle \Omega_a({\cal E}) | \lambda;J0 \rangle|^2 \, \nu_{J}^2 \, \delta_{\mu \nu} \,
\delta_{\lambda \lambda^{\prime}}   \nonumber \\
& = P_J({\cal E}) \, \nu_{J}^2 \,  \delta_{\mu \nu} 
\end{align}
in terms of $P_J({\cal E})$, the probability for the incident channel to enter the bound state manifold with total angular momentum $J$, as defined in Eq.~\eqref{PJ}.

\section{Field evolution of the coefficients $x_J$}

\label{sec:FIELDEVOL}

Based on the preceding, we can now define the approximate electric-field-dependent 
indirect absorption coefficient
\begin{equation}
x_{J}({\cal E}) = \frac{\pi^2 \,  \, \nu^2}{d_J}P_J({\cal E}) 
\label{xofJinE}
\end{equation}
where as a final approximation we have assumed the variance $\nu_J^2 = \nu^2$ is independent of the total angular momentum.
As we consider the ultracold quantum regime for ground rotational state molecules $n_1=n_2=0$,
$J$ can only take two values at zero field: $J=0$ ($J=1$) for identical and indistinguishable bosons (fermions),
due to the $s$-wave $l=0$ ($p$-wave $l=1$) orbital angular momenta.
We note that this value $J_0$ is the unique and well defined value of $J$ when the electric field is zero.
Then
\begin{equation}
x_{J}({\cal E}) \, \bigg|_{{\cal E}=0} = \frac{\pi^2 \, \nu^2}{d_{J_0}} \, \delta_{J,J_0} \equiv x_{J_0}(0)
\label{xofJinEzero}
\end{equation}
as $P_J({\cal E}) \underset{{\cal E} \to 0}{\longrightarrow} \delta_{J,J_0}$,
as can be seen in Eq.~\eqref{P0-B}, Eq.~\eqref{P1-F} and
Fig.~\eqref{FIG-PROBA-ADIM}.
Using Eq.~\eqref{xofJinEzero} in Eq.~\eqref{xofJinE},
we get
\begin{equation}
x_J({\cal E}) = x_{J_0}(0)  \, \frac{d_{J_0}}{d_{J}} \, P_J({\cal E}) =
x_{J_0}(0)  \, \frac{\rho_J}{\rho_{J_0}} \, P_J({\cal E})  .
\end{equation}
This is our main result.  It provides insight into the way in which complexes 
of different total angular momentum $J$ are populated, 
by extrapolating in a regular way from the result at zero field.
The coefficient $x_{J_0}(0)$ is extracted from experiments
performed at zero electric fields. 
For example,
the best fit values are 0.5~\cite{Bai_PRA_100_012705_2019,Ye_SA_4_eaaq0083_2018} for NaRb
and 0.26~\cite{Gregory_NC_10_3104_2019} for RbCs.
We now focus explicitly on the two possible cases
of bosonic and fermionic systems.

\begin{figure}[h]
\begin{center}
\includegraphics*[width=8cm, trim=0cm 0cm 0cm 0cm]{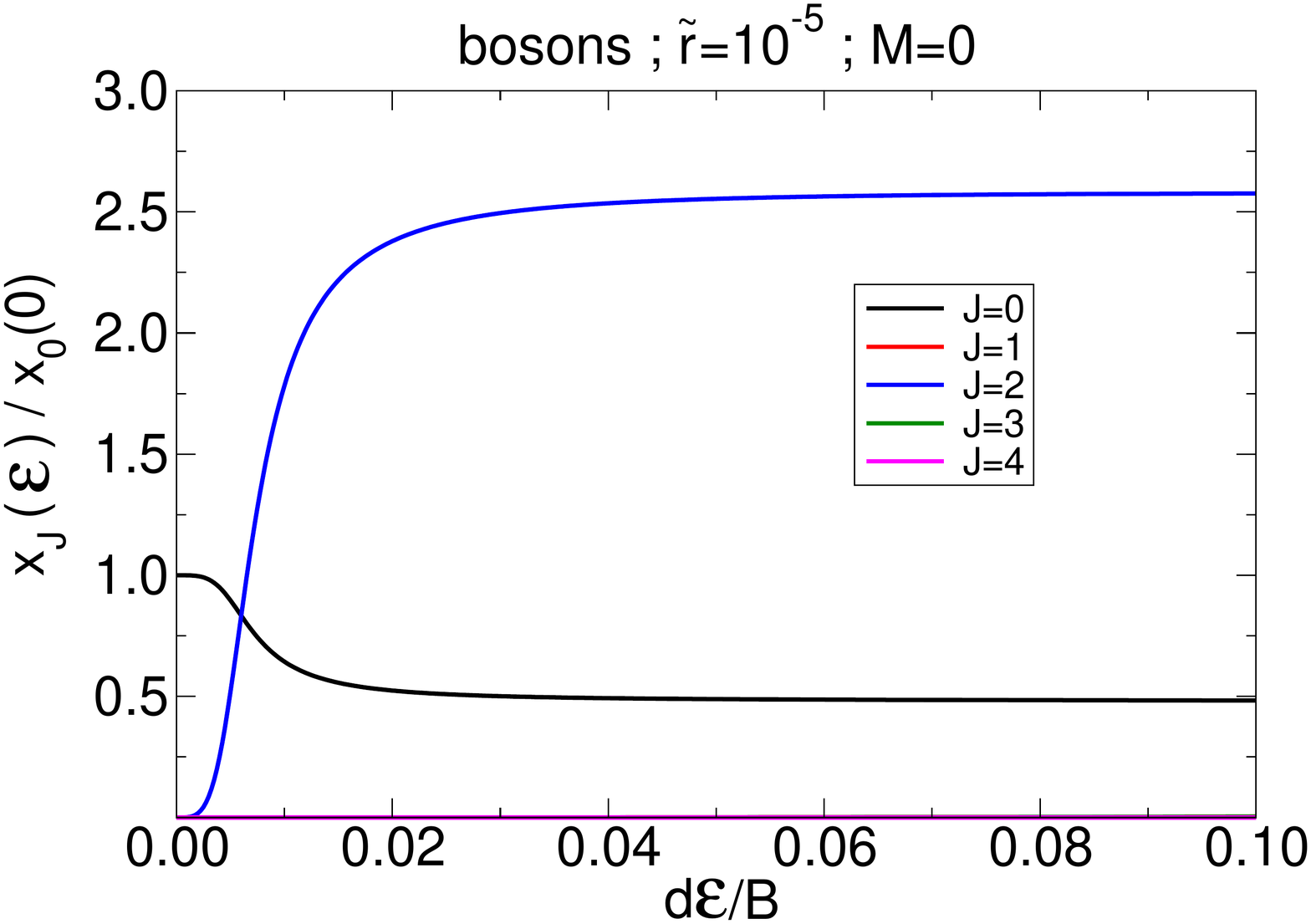}
\caption{The quantity $x_J({\cal E}) / x_0(0)$ as a function of $d {\cal E} /  B$ at a value of $\tilde{r}=10^{-5}$,
for the different allowed values of $J$ with $M=0$, for the bosonic system.
}
\label{FIG-XOFJ-BOS}
\end{center}
\end{figure}

\begin{figure}[h]
\begin{center}
\includegraphics*[width=8cm, trim=0cm 0cm 0cm 0cm]{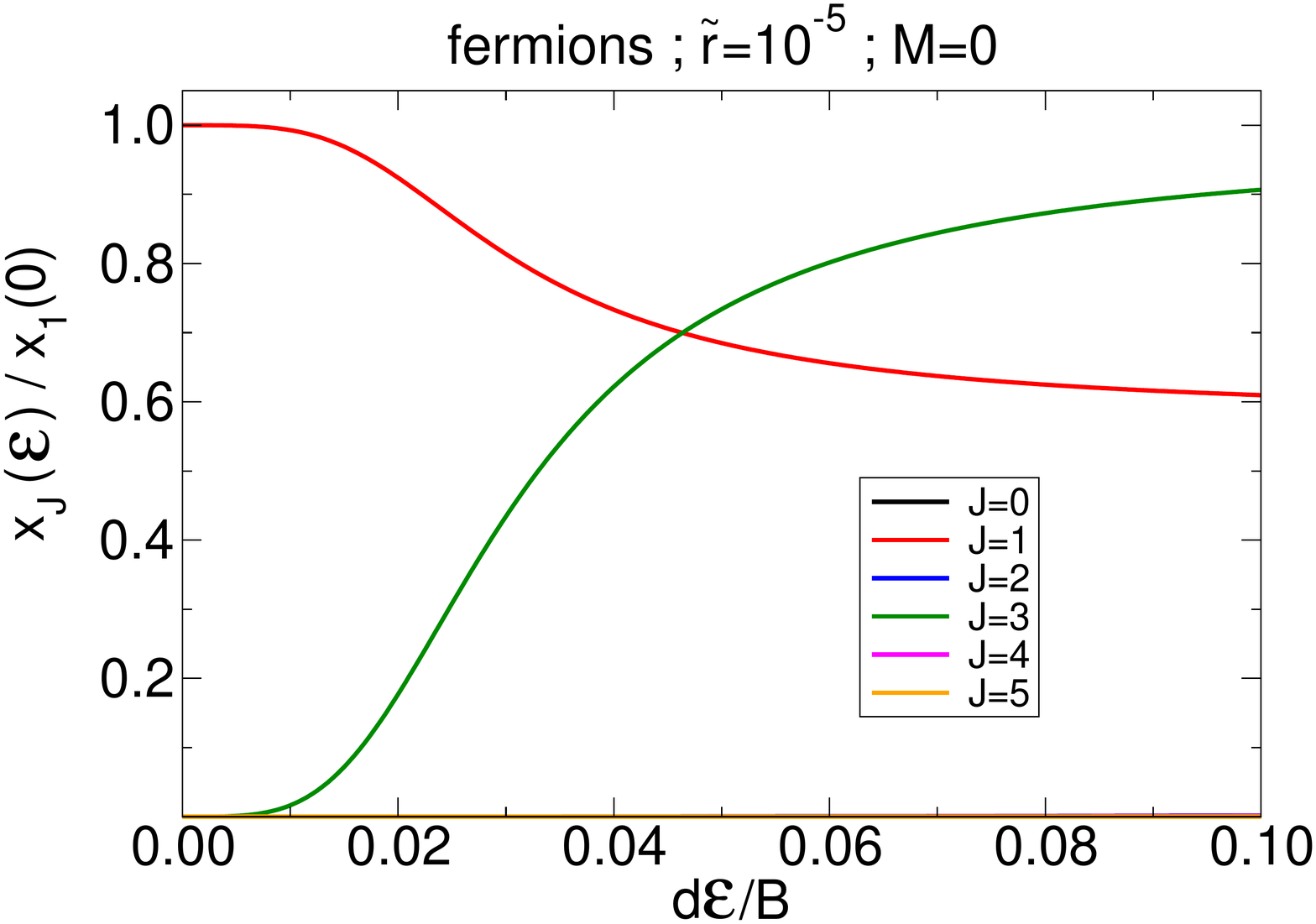}
\caption{Same as Fig.~\ref{FIG-XOFJ-BOS} but for the fermionic system.
}
\label{FIG-XOFJ-FER}
\end{center}
\end{figure}

\subsubsection{The $x$ coefficient for bosons}

In this case, $J_0=0$. Noting that $\rho_J/\rho_{J_0} = 2J+1$ for a given $M$
as detailed in Eq.~36 of~\cite{Christianen_PRA_100_032708_2019}
or as also found by looking at the ratio of the density-of-states
in~\cite{Mayle_PRA_87_012709_2013}, we find finally
\begin{equation}
x_J({\cal E}) = x_0(0)  \, (2J+1) \, P_J({\cal E}) .
\label{xJ-B}
\end{equation}
A priori, we don't sum these coefficients
as the $J$ components are not good quantum numbers. Instead, we have to combine the different $x_J$ using similar arguments
given in Eq. 55 of~\cite{Croft_PRA_102_033306_2020}.
As we are interested in the change of the $x$ coefficient for small values of the electric field,
we can simplify the resulting combination.
By looking at Fig.~\ref{FIG-PROBA-ADIM} for the bosonic case it is clear that
we can focus on the $J=0$ and $J=2$ components at small electric fields as they are
the dominant terms.
The total $x$ coefficient is thus given by
\begin{equation}
x({\cal E}) = \frac{x_0({\cal E}) + x_2({\cal E})}{1 + x_0({\cal E}) \, x_2({\cal E})} = \frac{x_0(0) \,  (P_0({\cal E}) + 5 \, P_2({\cal E}))}{1 + 5 \, [x_0(0)]^2 \, P_0({\cal E}) \, P_2({\cal E})}. \nonumber \\
\label{x-B}
\end{equation}

\subsubsection{The $x$ coefficient for fermions}

In this case, $J_0=1$.
Noting that $\rho_J/\rho_1 = (\rho_J/\rho_0)  (\rho_0/\rho_1)  = (2J+1) (1/3)$,  we find finally
\begin{equation}
x_J({\cal E}) = x_1(0)  \, \bigg( \frac{2J+1}{3} \bigg) \, P_J({\cal E}).
\label{xJ-F}
\end{equation}
Using a parallel argument to above, looking at Fig.~\ref{FIG-PROBA-ADIM} for the fermionic case it is clear that
we can focus on the $J=1$ and $J=3$ components at small electric fields as they are the dominant terms.
The total $x$ coefficient from the combination procedure is thus given by
\begin{equation}
x({\cal E}) = \frac{x_1({\cal E}) + x_3({\cal E})}{1 + x_1({\cal E}) \, x_3({\cal E})} = \frac{x_1(0) \,  (P_1({\cal E})+ 7/3 \, P_3({\cal E})) }{1 + 7/3 \, [x_1(0)]^2 \, P_1({\cal E}) \, P_3({\cal E})}. \nonumber \\
\label{x-F}
\end{equation}
\\

We plot the quantities $x_J({\cal E}) / x_0(0)$ from Eq.~\eqref{xJ-B} ($x_J({\cal E}) / x_1(0)$ 
from Eq.~\eqref{xJ-F}),
in Fig.~\ref{FIG-XOFJ-BOS} (Fig.~\ref{FIG-XOFJ-FER}),
at a value of $\tilde{r}=10^{-5}$,
for the different allowed values of $J$ with $M=0$. The curves for $J=0,2$ ($J=1,3$) are dominant
in this range of $d{\cal E}/B$ so that one barely sees the other components in the figures.
The trend of the curves are similar to the trend of the $P_J^{B,F}({\cal E})$ ones,
but they include now the $J$-dependent density-of-states.
Because of this interplay, $x_2({\cal E})/x_0(0)$ becomes five times larger than $x_0({\cal E})/x_0(0)$ for the bosons
while $x_3({\cal E})/x_1(0)$ and $x_1({\cal E})/x_1(0)$ are about similar magnitudes.

\subsubsection{The absorption probability at short-range}

Using Eq.~\eqref{x-B} and Eq.~\eqref{x-F} and as mentioned earlier, 
the corresponding short-range absorption probability 
due to the resonances in the tetramer complex region for both the bosonic and fermionic cases is given by
\begin{equation}
\bar{p}_\text{res} = \frac{4 \, x({\cal E})}{(1+x({\cal E}))^2} \equiv \bar{p}_\text{abs} .
\label{PSR}
\end{equation}
When this probability is unity, the system is said to be universal as it does not depend at all
on the second QDT parameter $s$, responsible for the scattering phase-shift 
at short-range \cite{Idziaszek_PRL_104_113202_2010}.
We can extend a somewhat arbitrary range of universality
for which the coefficient $s$ provides only very small changes 
to the scattering observables \cite{Idziaszek_PRA_82_020703_2010}.
One can chose for example the range of universality 
$0.95 \le \bar{p}_\text{res} \le 1$
with a corresponding range $0.64 \le x \le 1.57$.
Then, when the extracted field-dependent $x({\cal E})$ coefficient lies within this range, 
it is expected that the field-dependent $s({\cal E})$ coefficient 
plays an insignificant role in the dynamics.
\\

\begin{figure}[t]
\begin{center}
\includegraphics*[width=8cm, trim=0cm 0cm 0cm 0cm]{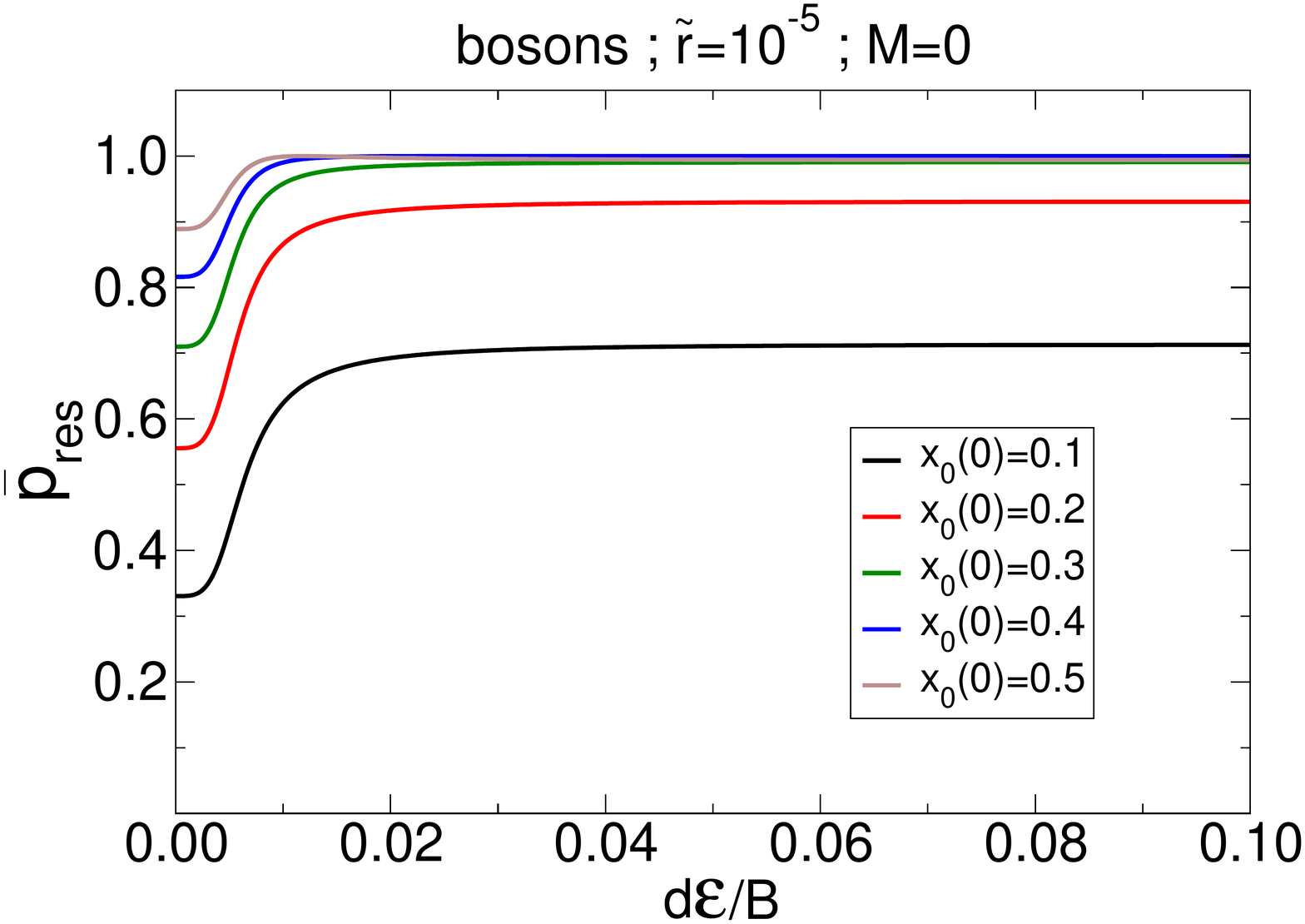} \\
\includegraphics*[width=8cm, trim=0cm 0cm 0cm 0cm]{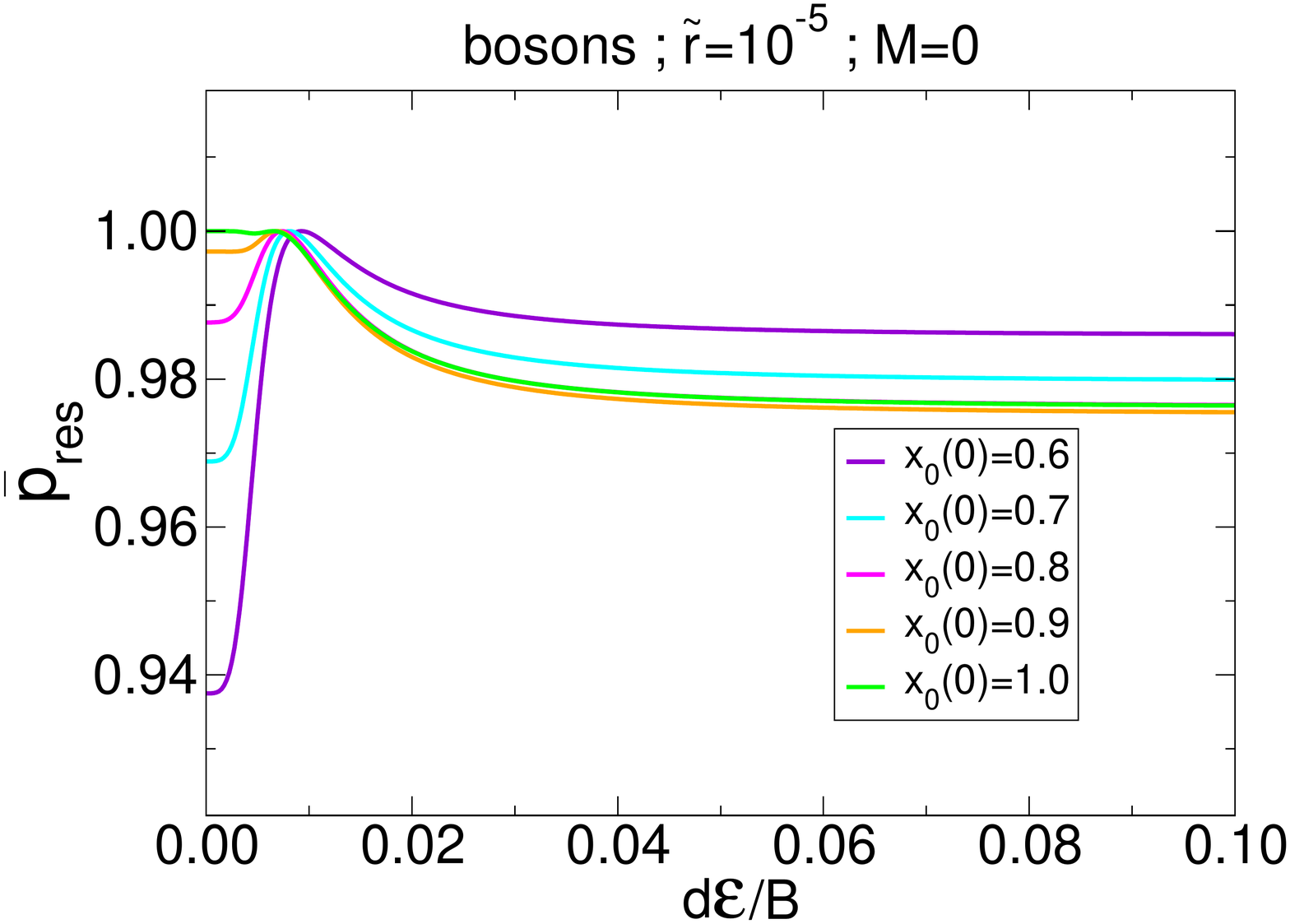}
\caption{Probability at short-range $\bar{p}_\mathrm{res}$
as a function of $d {\cal E} / B$ for the bosonic system
at a value of $\tilde{r}=10^{-5}$, for different zero-field values of $x_0(0)$.}
\label{FIG-PSR-B}
\end{center}
\end{figure}

\begin{figure}[t]
\begin{center}
\includegraphics*[width=8cm, trim=0cm 0cm 0cm 0cm]{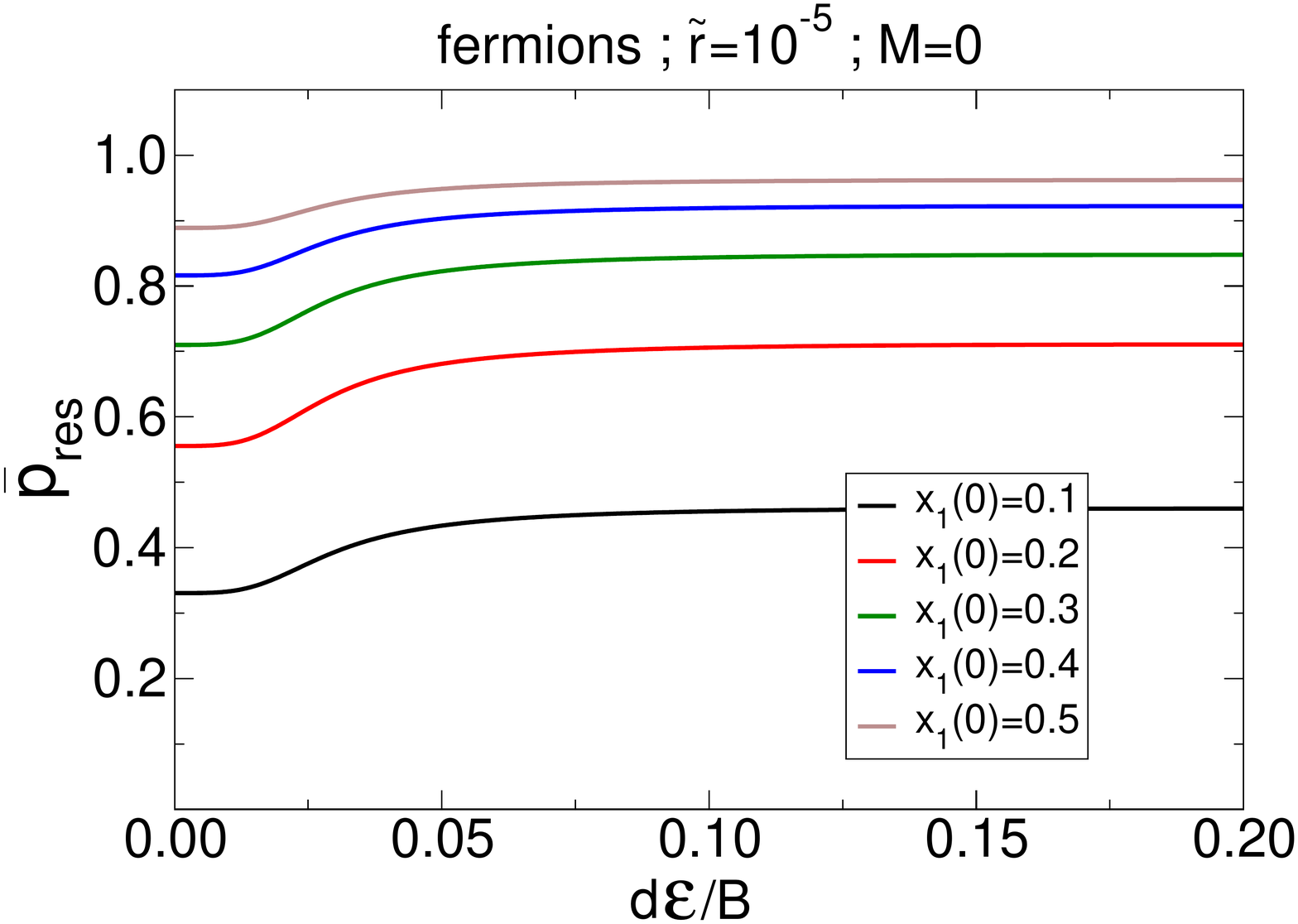} \\
\includegraphics*[width=8cm, trim=0cm 0cm 0cm 0cm]{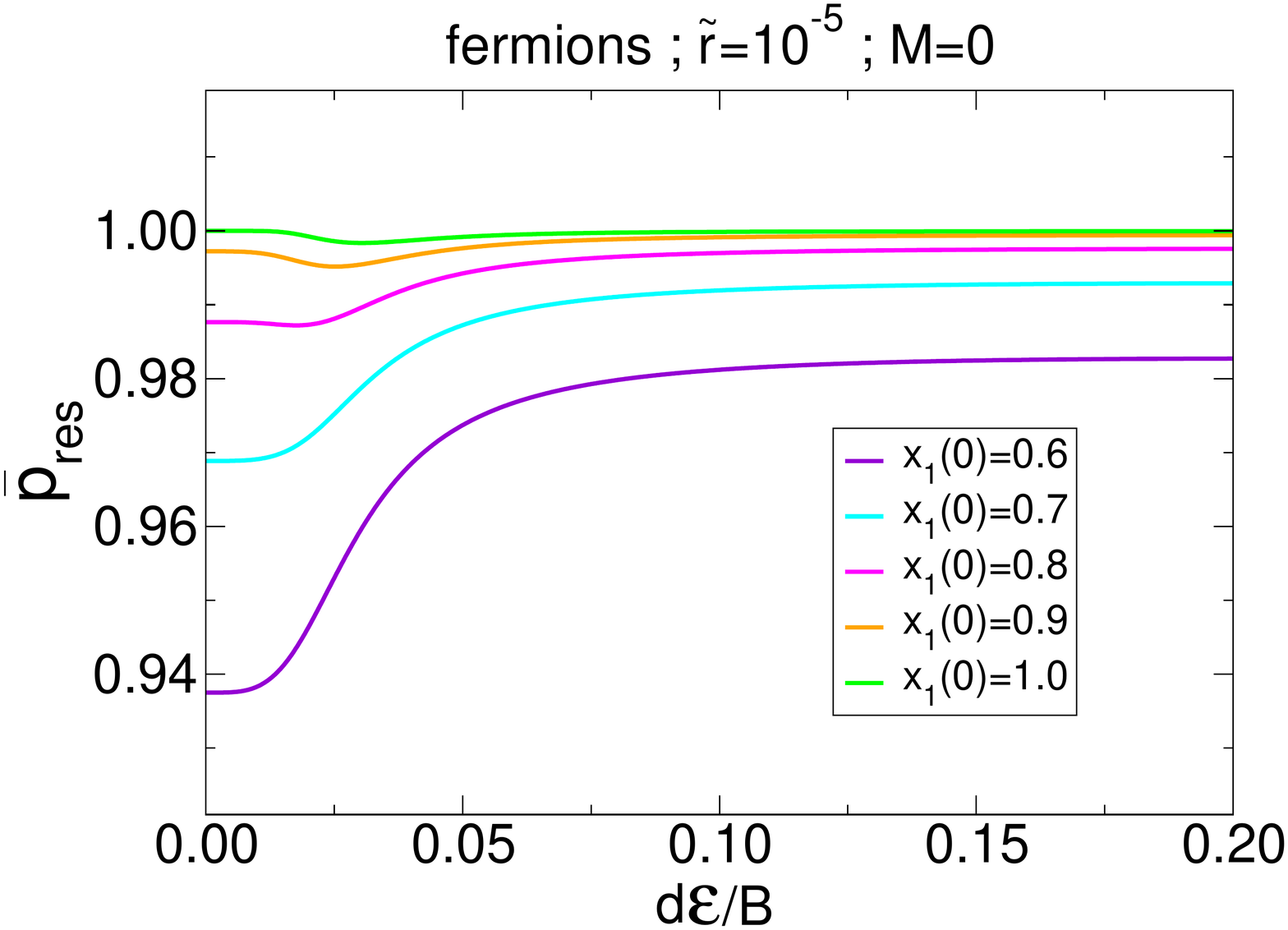}
\caption{Same as Fig.~\ref{FIG-PSR-B} but for the fermionic system,
for different zero-field values of $x_1(0)$.}
\label{FIG-PSR-F}
\end{center}
\end{figure}

In Fig.~\ref{FIG-PSR-B} (Fig.~\ref{FIG-PSR-F}),
we plot $\bar{p}_\mathrm{res}$
for different zero-field values of $x_0(0)$ ($x_1(0)$), for the bosonic (fermionic)
systems.
We emphasise that these are general and adimensional conclusions applicable
to any similar dipolar system.
In Sec.~\ref{APPLICATION} we will discuss their application to current systems
of experimental interest.

For the bosonic system, for the values of $x_0(0)=0.1,0.2$,
one can see that the zero-field behaviour is non-universal
with $\bar{p}_\mathrm{res} < 0.6$.
In-field, $\bar{p}_\mathrm{res}$ increases
and reaches values $\bar{p}_\mathrm{res} \simeq 0.9$ or smaller, still
not considered as the universal regime.
For the values $x_0(0)=0.3,0.4,0.5,0.6$ at zero field,
the behaviour could be qualified as not yet universal with $\bar{p}_\mathrm{res} < 0.95$.
In-field however, $\bar{p}_\mathrm{res}$ reaches values near unity and the behaviour is universal.
For the even larger values $x_0(0) \ge 0.7$, $\bar{p}_\mathrm{res} \ge 0.95$ for zero field
and $\bar{p}_\mathrm{res} > 0.97$ in-field, the behaviour is then universal in both cases.
Note that when $\bar{p}_\mathrm{res}$ starts very high, it may diminish slightly as the field is turned on.  This is due to the non-monotonic dependence of $\bar{p}_\mathrm{res}$ on $x$ in Eq.~\eqref{PSR}.  Nevertheless, the trend is clear: those collisions that are not universal in zero field tend to become more universal when the field is applied, while those that start universal in zero field remain so.  

A similar situation is seen for fermionic molecules. 
For the values of $x_1(0)=0.1,0.2,0.3,0.4$,
one can see that both the zero- and in-field behaviour are non-universal
with $\bar{p}_\mathrm{res} \le 0.9$.
For $x_1(0)=0.5, 0.6$, $\bar{p}_\mathrm{res} < 0.95$ at zero field,
not yet considered as the universal regime.
But in-field, $\bar{p}_\mathrm{res} > 0.95$ and the universal regime is reached.
Finally for $x_1(0) \ge 0.7$,
$\bar{p}_\mathrm{res} \ge 0.95$ for zero field
and $\bar{p}_\mathrm{res} > 0.98$ in-field. Then the zero and in-field behaviours, as for bosons,
become universal.

\section{Application to molecules of experimental interest}

\label{APPLICATION}

\subsection{Bosonic molecules}

\begin{figure}[h]
\begin{center}
\includegraphics*[width=8cm, trim=0cm 0cm 0cm 0cm]{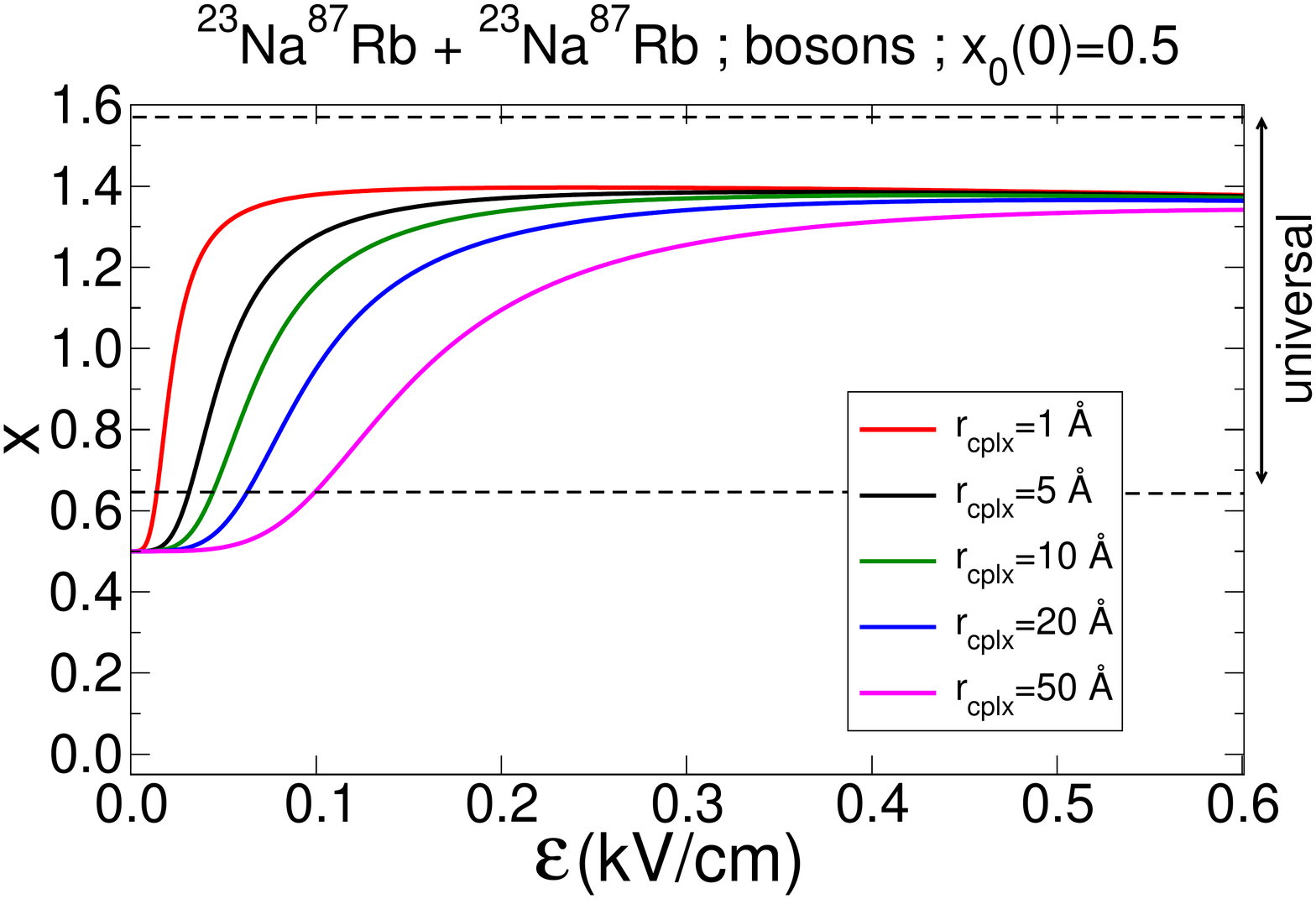}
\caption{Absorption coefficient at short-range $x$
as a function of ${\cal E}$ for the bosonic $^{23}$Na$^{87}$Rb +$^{23}$Na$^{87}$Rb system
at a zero-field value $x_0(0) = 0.5$ extracted
from experimental observations~\cite{Bai_PRA_100_012705_2019,Ye_SA_4_eaaq0083_2018}.
The red, black, green, blue, pink curves correspond to a characteristic position
$r = r_\text{cplx} = 1,5,10,20,50$ {\r{A}} where the tetramer complex stands.
The range $0.64 \le x \le 1.57$, where the corresponding short-range probability 
$0.95 \le \bar{p}_\text{res} \le 1$ is considered universal,
is indicated as dashed lines.}\label{FIG-PROBA-NARB}
\end{center}
\end{figure}

\begin{figure}[h]
\begin{center}
\includegraphics*[width=8cm, trim=0cm 0cm 0cm 0cm]{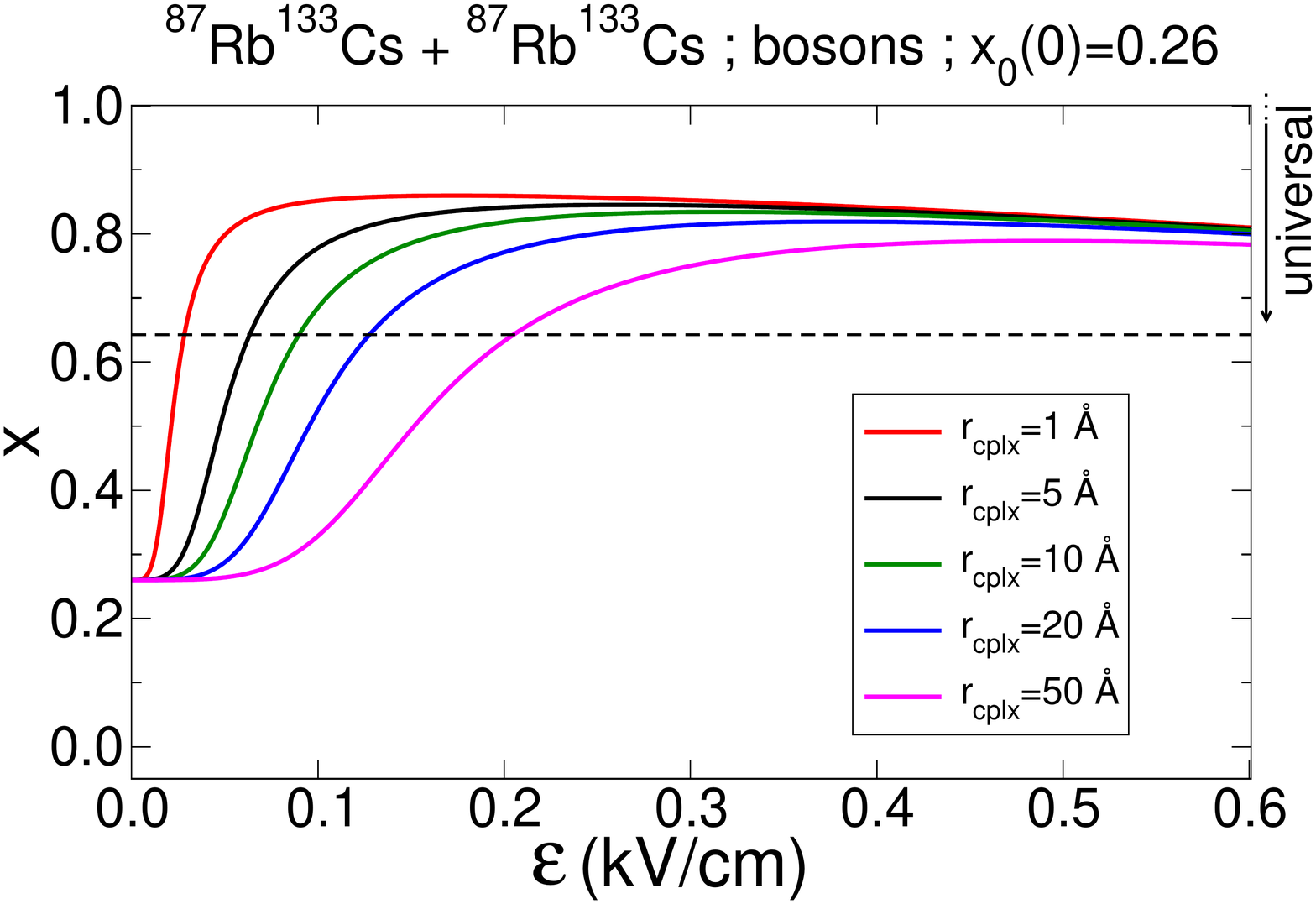}
\caption{Same as Fig.~\ref{FIG-PROBA-NARB} but for the bosonic $^{87}$Rb$^{133}$Cs + $^{87}$Rb$^{133}$Cs system
at a zero-field value $x_0(0) = 0.26$ extracted
from experimental observations~\cite{Gregory_NC_10_3104_2019}.}
\label{FIG-PROBA-RBCS}
\end{center}
\end{figure}

We turn to two bosonic systems of current experimental interest:
collisions of $^{23}$Na$^{87}$Rb +$^{23}$Na$^{87}$Rb~\cite{Ye_SA_4_eaaq0083_2018}
and of $^{87}$Rb$^{133}$Cs + $^{87}$Rb$^{133}$Cs~\cite{Gregory_NC_10_3104_2019},
for which zero-field values of $x_0(0)$
have been extracted from experimental observations.
We plot in Fig.~\ref{FIG-PROBA-NARB} and in Fig.~\ref{FIG-PROBA-RBCS}
the value of $x$ as a function of the electric field
for intial values $x_0(0) = 0.5$~\cite{Bai_PRA_100_012705_2019,Ye_SA_4_eaaq0083_2018}
for NaRb and $x_0(0) = 0.26$~\cite{Gregory_NC_10_3104_2019}
for RbCs.
This is done at different positions $r =  r_\text{cplx} =  1, 5, 10, 20, 50$ {\r{A}}.
The typical characteristic size of an alkali tetramer complex in its ground state
is around
$r_\text{cplx} \simeq$ 5--10~{\r{A}}~\cite{Byrd_PRA_82_010502_2010,Byrd_JCP_136_014306_2012,
Christianen_JCP_150_064106_2019,Yang_JPCL_11_2605_2020,
Klos_arXiv_2104_01625_2021}
so that the black and green curves correspond to a realistic estimation (within the scope of the model)
of what could be the value of $x$.
For indication, the limits $0.64 \le x \le 1.57$
where the universal regime $0.95 \le \bar{p}_\text{res} \le 1$ is reached
are displayed as dashed lines.
From these curves,
it is estimated that the NaRb and RbCs systems go from non-universal behaviour
at zero field to universal behaviour
in-field, within a small range of electric field,
from 0 to around 50~V/cm for NaRb and from 0 to around 100~V/cm for RbCs
(this is taken more or less when the black or green curves cross the dashed line).
This is a rapid change from non-universal to universal behaviour and future
experimental investigations could eventually probe this small
range of electric field, to observe this change of behaviour
by extracting the $x$ coefficient for each electric field.
For NaRb, the estimate for the transition seems consistent with the
experimental observation~\cite{Guo_PRX_8_041044_2018}.
For RbCs, as there is not yet an experimental observation in an electric field,
the present model predicts that the system is universal for fields higher
than 100~V/cm, even though the zero field behaviour is non-universal~\cite{Gregory_NC_10_3104_2019}.

\subsection{Fermionic molecules}

\begin{figure}[h]
\begin{center}
\includegraphics*[width=8cm, trim=0cm 0cm 0cm 0cm]{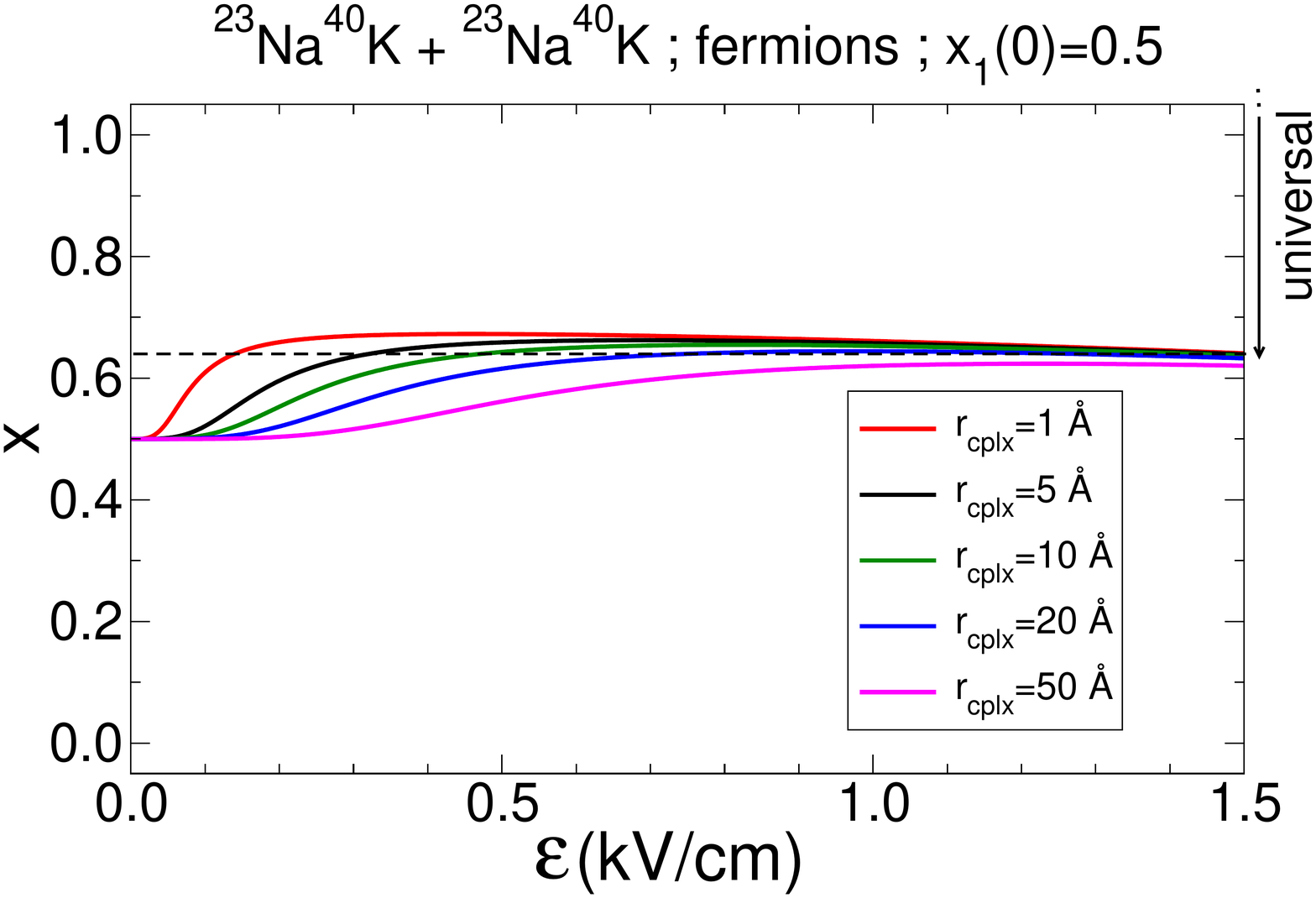}
\caption{Same as Fig.~\ref{FIG-PROBA-NARB} but for the fermionic $^{23}$Na$^{40}$K + $^{23}$Na$^{40}$K system
at a zero-field value $x_1(0) = 0.5$, an upper bound value of the coefficient that can be extrapolated from 
available experimental observations~\cite{Bause_PRR_3_033013_2021}.}
\label{FIG-PROBA-NAK}
\end{center}
\end{figure}

We also apply our model to a fermionic system 
$^{23}$Na$^{40}$K + $^{23}$Na$^{40}$K
of current experimental interest~\cite{Bause_PRR_3_033013_2021}.
Although it is not possible to extract a unique
zero-field value $x_1(0)$ from the experimental data,
a QDT analysis \cite{Idziaszek_PRL_104_113202_2010}
is still able to provide a bound on the QDT parameters $s$ and $x$.
To fit the experimental molecular loss slope of $\beta^\text{ls}/T \sim 13. \,10^{-11}$ cm$^3$/$\mu$K/s from \cite{Bause_PRR_3_033013_2021}, 
it is found that $x_1(0) \le 0.5$ for a range $1.6 \le s \le 2.7$,
as illustrated in Appendix~\ref{app:d}.

We use the upper limit $x_1(0) = 0.5$ as initial value and 
we plot in Fig.~\ref{FIG-PROBA-NAK} 
the value of $x$ as a function of the electric field.
The limit of the universal range is reached for a field of $\sim$ 300~V/cm when we choose 
$r_\text{cplx}= 5$~{\r{A}} and this would agree with the experimental results.
But in contrast with the bosonic systems studied above where the $x$ coefficients 
are well within the range of universality, the fermionic NaK system
remains at the limit of universality. Recall that we choose a 
lower universal limit of $\bar{p}_\text{res} = 0.95$ and this is somewhat arbitrary.

Interestingly, the experimental data of fermionic
NaK (see Fig.~3 in~\cite{Bause_PRR_3_033013_2021})
show a universal character in the sense that they do not present any oscillations
as a function of the electric field
(a feature that would have shown that deviation of universality is reached \cite{Idziaszek_PRA_82_020703_2010}).
But meanwhile, the overall background value of the loss rate coefficient
seems to be shifted compared to the theoretical universal prediction.
Both such conditions could be a feature that the system is within the limit of universality, as clearly 
displayed in Fig.~\ref{FIG-PROBA-NAK} here.

The fit to the experimental data just provides an upper limit to $x_1(0)$
and we took this upper limit as an example just above.
If this value turns out to be less than this upper limit, that is $x_1(0) < 0.5$,
the $x$ coefficient in the electric field will be below the range of universal character,
as can be seen in the upper panel of Fig.~\ref{FIG-PSR-F}.
For example, for the four curves where $x_1(0) \le 0.4$, $\bar{p}_\text{res} \le 0.92$,
which is deviating from universality. In that case, our model will not be able to explain why
the fermionic NaK system is near universal in an electric field in the experiment.
\\

Therefore, future experiments, where the small electric field range of these bosonic 
and fermionic systems is finely probed and the value $x({\cal E})$ is extracted with a high resolution,
will certainly shed light on the collision dynamics of such systems,
and should be able to validate the present model or not.

\subsection{Effect of the size of the complex}

Finally, we report in Fig.~\ref{FIG-PROBA-NARB}, Fig.~\ref{FIG-PROBA-RBCS} and Fig.~\ref{FIG-PROBA-NAK}, 
the effect of the tetramer complex range, by plotting the $x$ 
coefficient for additional values of $r_\text{cplx}= 1$, 20, 50~{\r{A}}.
While these values are a bit less realistic and more extreme, 
this illustrates the effect of a
smaller or bigger size of the complex on the range of the electric field in
which the systems go from non-universal to universal.
Typically, for a smaller (bigger) complex size, this range becomes
narrower (wider).
$r_\text{cplx}$ can eventually become an adjusting parameter in future experiments to fit 
the place where the sharp electric field feature of the universal character occurs.

\section{Conclusion}

\label{CONCLU}

We have proposed a theoretical model based on simple analytic formulas
to estimate the probability of absorption at short-range for dipolar collisions
in electric fields due to complex formation.
The model computes the amount of a $J$ component in the total wavefunction of
a dimer-dimer system when an electric field is turned on,
assuming that only the long-range physics is responsible for this $J$-mixing, 
given the small electric fields at which it occurs.
This is combined with the density-of-states of the tetramer complex
with a particular $J$ to give the QDT $x$ coefficient that determines the
probability of absorption due to complex formation at short-range,
and thus determines the scattering observables,
such as the cross sections and rate coefficients.
We treated both bosonic and fermionic cases
and applied the model to three systems of experimental interest.
This model shows that even though a system is non-universal in the absence of
an electric field, it can be universal as soon as a small electric field is
applied.
The range of electric fields over which this transition occurs is qualitatively related,
within this model, to the physical estimate of the size of the tetramer complex.
Future experiments on non-reactive ultracold molecular collisions in electric fields 
would then be important to validate this
model and to explain this change of universal character in collision of
ultracold non-reactive molecules.

Beyond the specific assumptions employed in the model (notably that the molecules make a choice about which total-$J$ collision complex to enter upon reaching a somewhat arbitrary intermolecular distance),
lies the clear qualitative notion that the rates of complex formation should depend on tunable external parameters, the electric field in this case.  We are therefore proposing here a kind of spectroscopy of the complex, where its properties and its coupling to the initial molecular channels can be varied and studied under controlled conditions. More detailed theoretical investigations will of course need to be performed to understand the outcomes of such spectroscopy. Still, this indirect probe may provide valuable insights into the few-body physics involved, in cases where the energy levels of the complex are too dense to resolve explicitly by conventional means.

\section*{Acknowledgments}

G. Q. acknowledges funding from the FEW2MANY-SHIELD Project No. ANR-17-CE30-0015 from Agence Nationale de la Recherche.
J. F. E. C. gratefully acknowledges support from the Dodd-Walls Centre for Photonic and Quantum Technologies.
J. L. B. acknowledges that this material is based on work supported by the National Science Foundation under grant number  1806971.

\bibliography{../../../BIBLIOGRAPHY/bibliography}

\begin{thebibliography}{37}
\expandafter\ifx\csname natexlab\endcsname\relax\def\natexlab#1{#1}\fi
\expandafter\ifx\csname bibnamefont\endcsname\relax
  \def\bibnamefont#1{#1}\fi
\expandafter\ifx\csname bibfnamefont\endcsname\relax
  \def\bibfnamefont#1{#1}\fi
\expandafter\ifx\csname citenamefont\endcsname\relax
  \def\citenamefont#1{#1}\fi
\expandafter\ifx\csname url\endcsname\relax
  \def\url#1{\texttt{#1}}\fi
\expandafter\ifx\csname urlprefix\endcsname\relax\def\urlprefix{URL }\fi
\providecommand{\bibinfo}[2]{#2}
\providecommand{\eprint}[2][]{\url{#2}}

\bibitem[{\citenamefont{Guo et~al.}(2016)\citenamefont{Guo, Zhu, Lu, Ye, Wang,
  Vexiau, Bouloufa-Maafa, Qu\'em\'ener, Dulieu, and
  Wang}}]{Guo_PRL_116_205303_2016}
\bibinfo{author}{\bibfnamefont{M.}~\bibnamefont{Guo}},
  \bibinfo{author}{\bibfnamefont{B.}~\bibnamefont{Zhu}},
  \bibinfo{author}{\bibfnamefont{B.}~\bibnamefont{Lu}},
  \bibinfo{author}{\bibfnamefont{X.}~\bibnamefont{Ye}},
  \bibinfo{author}{\bibfnamefont{F.}~\bibnamefont{Wang}},
  \bibinfo{author}{\bibfnamefont{R.}~\bibnamefont{Vexiau}},
  \bibinfo{author}{\bibfnamefont{N.}~\bibnamefont{Bouloufa-Maafa}},
  \bibinfo{author}{\bibfnamefont{G.}~\bibnamefont{Qu\'em\'ener}},
  \bibinfo{author}{\bibfnamefont{O.}~\bibnamefont{Dulieu}}, \bibnamefont{and}
  \bibinfo{author}{\bibfnamefont{D.}~\bibnamefont{Wang}},
  \bibinfo{journal}{Phys. Rev. Lett.} \textbf{\bibinfo{volume}{116}},
  \bibinfo{pages}{205303} (\bibinfo{year}{2016}).

\bibitem[{\citenamefont{Ye et~al.}(2018)\citenamefont{Ye, Guo,
  Gonz{\'a}lez-Mart{\'\i}nez, Qu{\'e}m{\'e}ner, and
  Wang}}]{Ye_SA_4_eaaq0083_2018}
\bibinfo{author}{\bibfnamefont{X.}~\bibnamefont{Ye}},
  \bibinfo{author}{\bibfnamefont{M.}~\bibnamefont{Guo}},
  \bibinfo{author}{\bibfnamefont{M.~L.}
  \bibnamefont{Gonz{\'a}lez-Mart{\'\i}nez}},
  \bibinfo{author}{\bibfnamefont{G.}~\bibnamefont{Qu{\'e}m{\'e}ner}},
  \bibnamefont{and} \bibinfo{author}{\bibfnamefont{D.}~\bibnamefont{Wang}},
  \bibinfo{journal}{Science Advances} \textbf{\bibinfo{volume}{4}},
  \bibinfo{pages}{eaaq0083} (\bibinfo{year}{2018}).

\bibitem[{\citenamefont{Guo et~al.}(2018)\citenamefont{Guo, Ye, He,
  Gonz\'alez-Mart\'{\i}nez, Vexiau, Qu\'em\'ener, and
  Wang}}]{Guo_PRX_8_041044_2018}
\bibinfo{author}{\bibfnamefont{M.}~\bibnamefont{Guo}},
  \bibinfo{author}{\bibfnamefont{X.}~\bibnamefont{Ye}},
  \bibinfo{author}{\bibfnamefont{J.}~\bibnamefont{He}},
  \bibinfo{author}{\bibfnamefont{M.~L.}
  \bibnamefont{Gonz\'alez-Mart\'{\i}nez}},
  \bibinfo{author}{\bibfnamefont{R.}~\bibnamefont{Vexiau}},
  \bibinfo{author}{\bibfnamefont{G.}~\bibnamefont{Qu\'em\'ener}},
  \bibnamefont{and} \bibinfo{author}{\bibfnamefont{D.}~\bibnamefont{Wang}},
  \bibinfo{journal}{Phys. Rev. X} \textbf{\bibinfo{volume}{8}},
  \bibinfo{pages}{041044} (\bibinfo{year}{2018}).

\bibitem[{\citenamefont{Takekoshi et~al.}(2014)\citenamefont{Takekoshi,
  Reichs\"ollner, Schindewolf, Hutson, Le~Sueur, Dulieu, Ferlaino, Grimm, and
  N\"agerl}}]{Takekoshi_PRL_113_205301_2014}
\bibinfo{author}{\bibfnamefont{T.}~\bibnamefont{Takekoshi}},
  \bibinfo{author}{\bibfnamefont{L.}~\bibnamefont{Reichs\"ollner}},
  \bibinfo{author}{\bibfnamefont{A.}~\bibnamefont{Schindewolf}},
  \bibinfo{author}{\bibfnamefont{J.~M.} \bibnamefont{Hutson}},
  \bibinfo{author}{\bibfnamefont{C.~R.} \bibnamefont{Le~Sueur}},
  \bibinfo{author}{\bibfnamefont{O.}~\bibnamefont{Dulieu}},
  \bibinfo{author}{\bibfnamefont{F.}~\bibnamefont{Ferlaino}},
  \bibinfo{author}{\bibfnamefont{R.}~\bibnamefont{Grimm}}, \bibnamefont{and}
  \bibinfo{author}{\bibfnamefont{H.-C.} \bibnamefont{N\"agerl}},
  \bibinfo{journal}{Phys. Rev. Lett.} \textbf{\bibinfo{volume}{113}},
  \bibinfo{pages}{205301} (\bibinfo{year}{2014}).

\bibitem[{\citenamefont{Molony et~al.}(2014)\citenamefont{Molony, Gregory, Ji,
  Lu, K\"oppinger, Le~Sueur, Blackley, Hutson, and
  Cornish}}]{Molony_PRL_113_255301_2014}
\bibinfo{author}{\bibfnamefont{P.~K.} \bibnamefont{Molony}},
  \bibinfo{author}{\bibfnamefont{P.~D.} \bibnamefont{Gregory}},
  \bibinfo{author}{\bibfnamefont{Z.}~\bibnamefont{Ji}},
  \bibinfo{author}{\bibfnamefont{B.}~\bibnamefont{Lu}},
  \bibinfo{author}{\bibfnamefont{M.~P.} \bibnamefont{K\"oppinger}},
  \bibinfo{author}{\bibfnamefont{C.~R.} \bibnamefont{Le~Sueur}},
  \bibinfo{author}{\bibfnamefont{C.~L.} \bibnamefont{Blackley}},
  \bibinfo{author}{\bibfnamefont{J.~M.} \bibnamefont{Hutson}},
  \bibnamefont{and} \bibinfo{author}{\bibfnamefont{S.~L.}
  \bibnamefont{Cornish}}, \bibinfo{journal}{Phys. Rev. Lett.}
  \textbf{\bibinfo{volume}{113}}, \bibinfo{pages}{255301}
  (\bibinfo{year}{2014}).

\bibitem[{\citenamefont{{Gregory} et~al.}(2019)\citenamefont{{Gregory}, {Frye},
  {Blackmore}, {Bridge}, {Sawant}, {Hutson}, and
  {Cornish}}}]{Gregory_NC_10_3104_2019}
\bibinfo{author}{\bibfnamefont{P.~D.} \bibnamefont{{Gregory}}},
  \bibinfo{author}{\bibfnamefont{M.~D.} \bibnamefont{{Frye}}},
  \bibinfo{author}{\bibfnamefont{J.~A.} \bibnamefont{{Blackmore}}},
  \bibinfo{author}{\bibfnamefont{E.~M.} \bibnamefont{{Bridge}}},
  \bibinfo{author}{\bibfnamefont{R.}~\bibnamefont{{Sawant}}},
  \bibinfo{author}{\bibfnamefont{J.~M.} \bibnamefont{{Hutson}}},
  \bibnamefont{and} \bibinfo{author}{\bibfnamefont{S.~L.}
  \bibnamefont{{Cornish}}}, \bibinfo{journal}{Nat. Commun.}
  \textbf{\bibinfo{volume}{10}}, \bibinfo{pages}{3104} (\bibinfo{year}{2019}).

\bibitem[{\citenamefont{Voges et~al.}(2020)\citenamefont{Voges, Gersema,
  Meyer~zum Alten~Borgloh, Schulze, Hartmann, Zenesini, and
  Ospelkaus}}]{Voges_PRL_125_083401_2020}
\bibinfo{author}{\bibfnamefont{K.~K.} \bibnamefont{Voges}},
  \bibinfo{author}{\bibfnamefont{P.}~\bibnamefont{Gersema}},
  \bibinfo{author}{\bibfnamefont{M.}~\bibnamefont{Meyer~zum Alten~Borgloh}},
  \bibinfo{author}{\bibfnamefont{T.~A.} \bibnamefont{Schulze}},
  \bibinfo{author}{\bibfnamefont{T.}~\bibnamefont{Hartmann}},
  \bibinfo{author}{\bibfnamefont{A.}~\bibnamefont{Zenesini}}, \bibnamefont{and}
  \bibinfo{author}{\bibfnamefont{S.}~\bibnamefont{Ospelkaus}},
  \bibinfo{journal}{Phys. Rev. Lett.} \textbf{\bibinfo{volume}{125}},
  \bibinfo{pages}{083401} (\bibinfo{year}{2020}).

\bibitem[{\citenamefont{Park et~al.}(2015)\citenamefont{Park, Will, and
  Zwierlein}}]{Park_PRL_114_205302_2015}
\bibinfo{author}{\bibfnamefont{J.~W.} \bibnamefont{Park}},
  \bibinfo{author}{\bibfnamefont{S.~A.} \bibnamefont{Will}}, \bibnamefont{and}
  \bibinfo{author}{\bibfnamefont{M.~W.} \bibnamefont{Zwierlein}},
  \bibinfo{journal}{Phys. Rev. Lett.} \textbf{\bibinfo{volume}{114}},
  \bibinfo{pages}{205302} (\bibinfo{year}{2015}).

\bibitem[{\citenamefont{Bause et~al.}(2021)\citenamefont{Bause, Schindewolf,
  Tao, Duda, Chen, Qu\'em\'ener, Karman, Christianen, Bloch, and
  Luo}}]{Bause_PRR_3_033013_2021}
\bibinfo{author}{\bibfnamefont{R.}~\bibnamefont{Bause}},
  \bibinfo{author}{\bibfnamefont{A.}~\bibnamefont{Schindewolf}},
  \bibinfo{author}{\bibfnamefont{R.}~\bibnamefont{Tao}},
  \bibinfo{author}{\bibfnamefont{M.}~\bibnamefont{Duda}},
  \bibinfo{author}{\bibfnamefont{X.-Y.} \bibnamefont{Chen}},
  \bibinfo{author}{\bibfnamefont{G.}~\bibnamefont{Qu\'em\'ener}},
  \bibinfo{author}{\bibfnamefont{T.}~\bibnamefont{Karman}},
  \bibinfo{author}{\bibfnamefont{A.}~\bibnamefont{Christianen}},
  \bibinfo{author}{\bibfnamefont{I.}~\bibnamefont{Bloch}}, \bibnamefont{and}
  \bibinfo{author}{\bibfnamefont{X.-Y.} \bibnamefont{Luo}},
  \bibinfo{journal}{Phys. Rev. Research} \textbf{\bibinfo{volume}{3}},
  \bibinfo{pages}{033013} (\bibinfo{year}{2021}).

\bibitem[{\citenamefont{Mayle et~al.}(2013)\citenamefont{Mayle, Qu\'em\'ener,
  Ruzic, and Bohn}}]{Mayle_PRA_87_012709_2013}
\bibinfo{author}{\bibfnamefont{M.}~\bibnamefont{Mayle}},
  \bibinfo{author}{\bibfnamefont{G.}~\bibnamefont{Qu\'em\'ener}},
  \bibinfo{author}{\bibfnamefont{B.~P.} \bibnamefont{Ruzic}}, \bibnamefont{and}
  \bibinfo{author}{\bibfnamefont{J.~L.} \bibnamefont{Bohn}},
  \bibinfo{journal}{Phys. Rev. A} \textbf{\bibinfo{volume}{87}},
  \bibinfo{pages}{012709} (\bibinfo{year}{2013}).

\bibitem[{\citenamefont{Christianen
  et~al.}(2019{\natexlab{a}})\citenamefont{Christianen, Zwierlein, Groenenboom,
  and Karman}}]{Christianen_PRL_123_123402_2019}
\bibinfo{author}{\bibfnamefont{A.}~\bibnamefont{Christianen}},
  \bibinfo{author}{\bibfnamefont{M.~W.} \bibnamefont{Zwierlein}},
  \bibinfo{author}{\bibfnamefont{G.~C.} \bibnamefont{Groenenboom}},
  \bibnamefont{and} \bibinfo{author}{\bibfnamefont{T.}~\bibnamefont{Karman}},
  \bibinfo{journal}{Phys. Rev. Lett.} \textbf{\bibinfo{volume}{123}},
  \bibinfo{pages}{123402} (\bibinfo{year}{2019}{\natexlab{a}}).

\bibitem[{\citenamefont{{Klos} et~al.}(2021)\citenamefont{{Klos}, {Guan}, {Li},
  {Li}, {Tiesinga}, and {Kotochigova}}}]{Klos_arXiv_2104_01625_2021}
\bibinfo{author}{\bibfnamefont{J.}~\bibnamefont{{Klos}}},
  \bibinfo{author}{\bibfnamefont{Q.}~\bibnamefont{{Guan}}},
  \bibinfo{author}{\bibfnamefont{H.}~\bibnamefont{{Li}}},
  \bibinfo{author}{\bibfnamefont{M.}~\bibnamefont{{Li}}},
  \bibinfo{author}{\bibfnamefont{E.}~\bibnamefont{{Tiesinga}}},
  \bibnamefont{and}
  \bibinfo{author}{\bibfnamefont{S.}~\bibnamefont{{Kotochigova}}},
  \bibinfo{journal}{ArXiv e-prints} p. \bibinfo{pages}{2104.01625}
  (\bibinfo{year}{2021}).

\bibitem[{\citenamefont{Liu et~al.}(2020)\citenamefont{Liu, Hu, Nichols,
  Grimes, Karman, Guo, and Ni}}]{Liu_NP_16_1132_2020}
\bibinfo{author}{\bibfnamefont{Y.}~\bibnamefont{Liu}},
  \bibinfo{author}{\bibfnamefont{M.-G.} \bibnamefont{Hu}},
  \bibinfo{author}{\bibfnamefont{M.~A.} \bibnamefont{Nichols}},
  \bibinfo{author}{\bibfnamefont{D.~D.} \bibnamefont{Grimes}},
  \bibinfo{author}{\bibfnamefont{T.}~\bibnamefont{Karman}},
  \bibinfo{author}{\bibfnamefont{H.}~\bibnamefont{Guo}}, \bibnamefont{and}
  \bibinfo{author}{\bibfnamefont{K.-K.} \bibnamefont{Ni}},
  \bibinfo{journal}{Nat. Phys.} \textbf{\bibinfo{volume}{16}},
  \bibinfo{pages}{1132} (\bibinfo{year}{2020}).

\bibitem[{\citenamefont{Gregory et~al.}(2020)\citenamefont{Gregory, Blackmore,
  Bromley, and Cornish}}]{Gregory_PRL_124_163402_2020}
\bibinfo{author}{\bibfnamefont{P.~D.} \bibnamefont{Gregory}},
  \bibinfo{author}{\bibfnamefont{J.~A.} \bibnamefont{Blackmore}},
  \bibinfo{author}{\bibfnamefont{S.~L.} \bibnamefont{Bromley}},
  \bibnamefont{and} \bibinfo{author}{\bibfnamefont{S.~L.}
  \bibnamefont{Cornish}}, \bibinfo{journal}{Phys. Rev. Lett.}
  \textbf{\bibinfo{volume}{124}}, \bibinfo{pages}{163402}
  (\bibinfo{year}{2020}).

\bibitem[{\citenamefont{{Gersema} et~al.}(2021)\citenamefont{{Gersema},
  {Voges}, {Meyer zum Alten Borgloh}, {Koch}, {Hartmann}, {Zenesini},
  {Ospelkaus}, {Lin}, {He}, and {Wang}}}]{Gersema_arXiv_2103_00510_2021}
\bibinfo{author}{\bibfnamefont{P.}~\bibnamefont{{Gersema}}},
  \bibinfo{author}{\bibfnamefont{K.~K.} \bibnamefont{{Voges}}},
  \bibinfo{author}{\bibfnamefont{M.}~\bibnamefont{{Meyer zum Alten Borgloh}}},
  \bibinfo{author}{\bibfnamefont{L.}~\bibnamefont{{Koch}}},
  \bibinfo{author}{\bibfnamefont{T.}~\bibnamefont{{Hartmann}}},
  \bibinfo{author}{\bibfnamefont{A.}~\bibnamefont{{Zenesini}}},
  \bibinfo{author}{\bibfnamefont{S.}~\bibnamefont{{Ospelkaus}}},
  \bibinfo{author}{\bibfnamefont{J.}~\bibnamefont{{Lin}}},
  \bibinfo{author}{\bibfnamefont{J.}~\bibnamefont{{He}}}, \bibnamefont{and}
  \bibinfo{author}{\bibfnamefont{D.}~\bibnamefont{{Wang}}},
  \bibinfo{journal}{ArXiv e-prints} p. \bibinfo{pages}{2103.00510}
  (\bibinfo{year}{2021}).

\bibitem[{\citenamefont{Hu et~al.}(2019)\citenamefont{Hu, Liu, Grimes, Lin,
  Gheorghe, Vexiau, Bouloufa-Maafa, Dulieu, Rosenband, and
  Ni}}]{Hu_S_366_1111_2019}
\bibinfo{author}{\bibfnamefont{M.-G.} \bibnamefont{Hu}},
  \bibinfo{author}{\bibfnamefont{Y.}~\bibnamefont{Liu}},
  \bibinfo{author}{\bibfnamefont{D.~D.} \bibnamefont{Grimes}},
  \bibinfo{author}{\bibfnamefont{Y.-W.} \bibnamefont{Lin}},
  \bibinfo{author}{\bibfnamefont{A.~H.} \bibnamefont{Gheorghe}},
  \bibinfo{author}{\bibfnamefont{R.}~\bibnamefont{Vexiau}},
  \bibinfo{author}{\bibfnamefont{N.}~\bibnamefont{Bouloufa-Maafa}},
  \bibinfo{author}{\bibfnamefont{O.}~\bibnamefont{Dulieu}},
  \bibinfo{author}{\bibfnamefont{T.}~\bibnamefont{Rosenband}},
  \bibnamefont{and} \bibinfo{author}{\bibfnamefont{K.-K.} \bibnamefont{Ni}},
  \bibinfo{journal}{Science} \textbf{\bibinfo{volume}{366}},
  \bibinfo{pages}{1111} (\bibinfo{year}{2019}).

\bibitem[{\citenamefont{Bai et~al.}(2019)\citenamefont{Bai, Li, Wang, and
  Cong}}]{Bai_PRA_100_012705_2019}
\bibinfo{author}{\bibfnamefont{Y.-P.} \bibnamefont{Bai}},
  \bibinfo{author}{\bibfnamefont{J.-L.} \bibnamefont{Li}},
  \bibinfo{author}{\bibfnamefont{G.-R.} \bibnamefont{Wang}}, \bibnamefont{and}
  \bibinfo{author}{\bibfnamefont{S.-L.} \bibnamefont{Cong}},
  \bibinfo{journal}{Phys. Rev. A} \textbf{\bibinfo{volume}{100}},
  \bibinfo{pages}{012705} (\bibinfo{year}{2019}).

\bibitem[{\citenamefont{Ospelkaus et~al.}(2010)\citenamefont{Ospelkaus, Ni,
  Wang, de~Miranda, Neyenhuis, Qu{\'e}m{\'e}ner, Julienne, Bohn, Jin, and
  Ye}}]{Ospelkaus_S_327_853_2010}
\bibinfo{author}{\bibfnamefont{S.}~\bibnamefont{Ospelkaus}},
  \bibinfo{author}{\bibfnamefont{K.-K.} \bibnamefont{Ni}},
  \bibinfo{author}{\bibfnamefont{D.}~\bibnamefont{Wang}},
  \bibinfo{author}{\bibfnamefont{M.~H.~G.} \bibnamefont{de~Miranda}},
  \bibinfo{author}{\bibfnamefont{B.}~\bibnamefont{Neyenhuis}},
  \bibinfo{author}{\bibfnamefont{G.}~\bibnamefont{Qu{\'e}m{\'e}ner}},
  \bibinfo{author}{\bibfnamefont{P.~S.} \bibnamefont{Julienne}},
  \bibinfo{author}{\bibfnamefont{J.~L.} \bibnamefont{Bohn}},
  \bibinfo{author}{\bibfnamefont{D.~S.} \bibnamefont{Jin}}, \bibnamefont{and}
  \bibinfo{author}{\bibfnamefont{J.}~\bibnamefont{Ye}},
  \bibinfo{journal}{Science} \textbf{\bibinfo{volume}{327}},
  \bibinfo{pages}{853} (\bibinfo{year}{2010}).

\bibitem[{\citenamefont{Ni et~al.}(2010)\citenamefont{Ni, Ospelkaus, Wang,
  Qu\'em\'ener, Neyenhuis, de~Miranda, Bohn, Jin, and Ye}}]{Ni_N_464_1324_2010}
\bibinfo{author}{\bibfnamefont{K.-K.} \bibnamefont{Ni}},
  \bibinfo{author}{\bibfnamefont{S.}~\bibnamefont{Ospelkaus}},
  \bibinfo{author}{\bibfnamefont{D.}~\bibnamefont{Wang}},
  \bibinfo{author}{\bibfnamefont{G.}~\bibnamefont{Qu\'em\'ener}},
  \bibinfo{author}{\bibfnamefont{B.}~\bibnamefont{Neyenhuis}},
  \bibinfo{author}{\bibfnamefont{M.~H.~G.} \bibnamefont{de~Miranda}},
  \bibinfo{author}{\bibfnamefont{J.~L.} \bibnamefont{Bohn}},
  \bibinfo{author}{\bibfnamefont{D.~S.} \bibnamefont{Jin}}, \bibnamefont{and}
  \bibinfo{author}{\bibfnamefont{J.}~\bibnamefont{Ye}},
  \bibinfo{journal}{Nature} \textbf{\bibinfo{volume}{464}},
  \bibinfo{pages}{1324} (\bibinfo{year}{2010}).

\bibitem[{\citenamefont{Qu\'em\'ener and
  Bohn}(2010)}]{Quemener_PRA_81_022702_2010}
\bibinfo{author}{\bibfnamefont{G.}~\bibnamefont{Qu\'em\'ener}}
  \bibnamefont{and} \bibinfo{author}{\bibfnamefont{J.~L.} \bibnamefont{Bohn}},
  \bibinfo{journal}{Phys. Rev. A} \textbf{\bibinfo{volume}{81}},
  \bibinfo{pages}{022702} (\bibinfo{year}{2010}).

\bibitem[{\citenamefont{Qu\'em\'ener et~al.}(2011)\citenamefont{Qu\'em\'ener,
  Bohn, Petrov, and Kotochigova}}]{Quemener_PRA_84_062703_2011}
\bibinfo{author}{\bibfnamefont{G.}~\bibnamefont{Qu\'em\'ener}},
  \bibinfo{author}{\bibfnamefont{J.~L.} \bibnamefont{Bohn}},
  \bibinfo{author}{\bibfnamefont{A.}~\bibnamefont{Petrov}}, \bibnamefont{and}
  \bibinfo{author}{\bibfnamefont{S.}~\bibnamefont{Kotochigova}},
  \bibinfo{journal}{Phys. Rev. A} \textbf{\bibinfo{volume}{84}},
  \bibinfo{pages}{062703} (\bibinfo{year}{2011}).

\bibitem[{\citenamefont{Micheli et~al.}(2010)\citenamefont{Micheli, Idziaszek,
  Pupillo, Baranov, Zoller, and Julienne}}]{Micheli_PRL_105_073202_2010}
\bibinfo{author}{\bibfnamefont{A.}~\bibnamefont{Micheli}},
  \bibinfo{author}{\bibfnamefont{Z.}~\bibnamefont{Idziaszek}},
  \bibinfo{author}{\bibfnamefont{G.}~\bibnamefont{Pupillo}},
  \bibinfo{author}{\bibfnamefont{M.~A.} \bibnamefont{Baranov}},
  \bibinfo{author}{\bibfnamefont{P.}~\bibnamefont{Zoller}}, \bibnamefont{and}
  \bibinfo{author}{\bibfnamefont{P.~S.} \bibnamefont{Julienne}},
  \bibinfo{journal}{Phys. Rev. Lett.} \textbf{\bibinfo{volume}{105}},
  \bibinfo{pages}{073202} (\bibinfo{year}{2010}).

\bibitem[{\citenamefont{Mayle et~al.}(2012)\citenamefont{Mayle, Ruzic, and
  Bohn}}]{Mayle_PRA_85_062712_2012}
\bibinfo{author}{\bibfnamefont{M.}~\bibnamefont{Mayle}},
  \bibinfo{author}{\bibfnamefont{B.~P.} \bibnamefont{Ruzic}}, \bibnamefont{and}
  \bibinfo{author}{\bibfnamefont{J.~L.} \bibnamefont{Bohn}},
  \bibinfo{journal}{Phys. Rev. A} \textbf{\bibinfo{volume}{85}},
  \bibinfo{pages}{062712} (\bibinfo{year}{2012}).

\bibitem[{\citenamefont{Croft et~al.}(2020)\citenamefont{Croft, Bohn, and
  Qu\'em\'ener}}]{Croft_PRA_102_033306_2020}
\bibinfo{author}{\bibfnamefont{J.~F.~E.} \bibnamefont{Croft}},
  \bibinfo{author}{\bibfnamefont{J.~L.} \bibnamefont{Bohn}}, \bibnamefont{and}
  \bibinfo{author}{\bibfnamefont{G.}~\bibnamefont{Qu\'em\'ener}},
  \bibinfo{journal}{Phys. Rev. A} \textbf{\bibinfo{volume}{102}},
  \bibinfo{pages}{033306} (\bibinfo{year}{2020}).

\bibitem[{\citenamefont{Idziaszek and
  Julienne}(2010)}]{Idziaszek_PRL_104_113202_2010}
\bibinfo{author}{\bibfnamefont{Z.}~\bibnamefont{Idziaszek}} \bibnamefont{and}
  \bibinfo{author}{\bibfnamefont{P.~S.} \bibnamefont{Julienne}},
  \bibinfo{journal}{Phys. Rev. Lett.} \textbf{\bibinfo{volume}{104}},
  \bibinfo{pages}{113202} (\bibinfo{year}{2010}).

\bibitem[{\citenamefont{Wang and
  Qu{\'e}m{\'e}ner}(2015)}]{Wang_NJP_17_035015_2015}
\bibinfo{author}{\bibfnamefont{G.}~\bibnamefont{Wang}} \bibnamefont{and}
  \bibinfo{author}{\bibfnamefont{G.}~\bibnamefont{Qu{\'e}m{\'e}ner}},
  \bibinfo{journal}{New J. Phys.} \textbf{\bibinfo{volume}{17}},
  \bibinfo{pages}{035015} (\bibinfo{year}{2015}).

\bibitem[{\citenamefont{Lepers et~al.}(2013)\citenamefont{Lepers, Vexiau,
  Aymar, Bouloufa-Maafa, and Dulieu}}]{Lepers_PRA_88_032709_2013}
\bibinfo{author}{\bibfnamefont{M.}~\bibnamefont{Lepers}},
  \bibinfo{author}{\bibfnamefont{R.}~\bibnamefont{Vexiau}},
  \bibinfo{author}{\bibfnamefont{M.}~\bibnamefont{Aymar}},
  \bibinfo{author}{\bibfnamefont{N.}~\bibnamefont{Bouloufa-Maafa}},
  \bibnamefont{and} \bibinfo{author}{\bibfnamefont{O.}~\bibnamefont{Dulieu}},
  \bibinfo{journal}{Phys. Rev. A} \textbf{\bibinfo{volume}{88}},
  \bibinfo{pages}{032709} (\bibinfo{year}{2013}).

\bibitem[{\citenamefont{Gao}(2008)}]{Gao_PRA_78_012702_2008}
\bibinfo{author}{\bibfnamefont{B.}~\bibnamefont{Gao}}, \bibinfo{journal}{Phys.
  Rev. A} \textbf{\bibinfo{volume}{78}}, \bibinfo{pages}{012702}
  (\bibinfo{year}{2008}).

\bibitem[{\citenamefont{Gonz\'alez-Mart\'{\i}nez
  et~al.}(2017)\citenamefont{Gonz\'alez-Mart\'{\i}nez, Bohn, and
  Qu\'em\'ener}}]{Gonzalez-Martinez_PRA_96_032718_2017}
\bibinfo{author}{\bibfnamefont{M.~L.} \bibnamefont{Gonz\'alez-Mart\'{\i}nez}},
  \bibinfo{author}{\bibfnamefont{J.~L.} \bibnamefont{Bohn}}, \bibnamefont{and}
  \bibinfo{author}{\bibfnamefont{G.}~\bibnamefont{Qu\'em\'ener}},
  \bibinfo{journal}{Phys. Rev. A} \textbf{\bibinfo{volume}{96}},
  \bibinfo{pages}{032718} (\bibinfo{year}{2017}).

\bibitem[{\citenamefont{Byrd et~al.}(2010)\citenamefont{Byrd, Montgomery, and
  C\^ot\'e}}]{Byrd_PRA_82_010502_2010}
\bibinfo{author}{\bibfnamefont{J.~N.} \bibnamefont{Byrd}},
  \bibinfo{author}{\bibfnamefont{J.~A.} \bibnamefont{Montgomery}},
  \bibnamefont{and} \bibinfo{author}{\bibfnamefont{R.}~\bibnamefont{C\^ot\'e}},
  \bibinfo{journal}{Phys. Rev. A} \textbf{\bibinfo{volume}{82}},
  \bibinfo{pages}{010502} (\bibinfo{year}{2010}).

\bibitem[{\citenamefont{Yang et~al.}(2020)\citenamefont{Yang, Zuo, Huang, Hu,
  Dawes, Xie, and Guo}}]{Yang_JPCL_11_2605_2020}
\bibinfo{author}{\bibfnamefont{D.}~\bibnamefont{Yang}},
  \bibinfo{author}{\bibfnamefont{J.}~\bibnamefont{Zuo}},
  \bibinfo{author}{\bibfnamefont{J.}~\bibnamefont{Huang}},
  \bibinfo{author}{\bibfnamefont{X.}~\bibnamefont{Hu}},
  \bibinfo{author}{\bibfnamefont{R.}~\bibnamefont{Dawes}},
  \bibinfo{author}{\bibfnamefont{D.}~\bibnamefont{Xie}}, \bibnamefont{and}
  \bibinfo{author}{\bibfnamefont{H.}~\bibnamefont{Guo}}, \bibinfo{journal}{J.
  Phys. Chem. Lett.} \textbf{\bibinfo{volume}{11}}, \bibinfo{pages}{2605}
  (\bibinfo{year}{2020}).

\bibitem[{\citenamefont{Christianen
  et~al.}(2019{\natexlab{b}})\citenamefont{Christianen, Karman,
  Vargas-Hernández, Groenenboom, and Krems}}]{Christianen_JCP_150_064106_2019}
\bibinfo{author}{\bibfnamefont{A.}~\bibnamefont{Christianen}},
  \bibinfo{author}{\bibfnamefont{T.}~\bibnamefont{Karman}},
  \bibinfo{author}{\bibfnamefont{R.~A.} \bibnamefont{Vargas-Hernández}},
  \bibinfo{author}{\bibfnamefont{G.~C.} \bibnamefont{Groenenboom}},
  \bibnamefont{and} \bibinfo{author}{\bibfnamefont{R.~V.} \bibnamefont{Krems}},
  \bibinfo{journal}{J. Chem. Phys.} \textbf{\bibinfo{volume}{150}},
  \bibinfo{pages}{064106} (\bibinfo{year}{2019}{\natexlab{b}}).

\bibitem[{\citenamefont{Christianen
  et~al.}(2019{\natexlab{c}})\citenamefont{Christianen, Karman, and
  Groenenboom}}]{Christianen_PRA_100_032708_2019}
\bibinfo{author}{\bibfnamefont{A.}~\bibnamefont{Christianen}},
  \bibinfo{author}{\bibfnamefont{T.}~\bibnamefont{Karman}}, \bibnamefont{and}
  \bibinfo{author}{\bibfnamefont{G.~C.} \bibnamefont{Groenenboom}},
  \bibinfo{journal}{Phys. Rev. A} \textbf{\bibinfo{volume}{100}},
  \bibinfo{pages}{032708} (\bibinfo{year}{2019}{\natexlab{c}}).

\bibitem[{\citenamefont{Croft et~al.}(2017)\citenamefont{Croft, Makrides, Li,
  Petrov, Kendrick, Balakrishnan, and Kotochigova}}]{Croft_NC_8_15897_2017}
\bibinfo{author}{\bibfnamefont{J.~F.~E.} \bibnamefont{Croft}},
  \bibinfo{author}{\bibfnamefont{C.}~\bibnamefont{Makrides}},
  \bibinfo{author}{\bibfnamefont{M.}~\bibnamefont{Li}},
  \bibinfo{author}{\bibfnamefont{A.}~\bibnamefont{Petrov}},
  \bibinfo{author}{\bibfnamefont{B.~K.} \bibnamefont{Kendrick}},
  \bibinfo{author}{\bibfnamefont{N.}~\bibnamefont{Balakrishnan}},
  \bibnamefont{and}
  \bibinfo{author}{\bibfnamefont{S.}~\bibnamefont{Kotochigova}},
  \bibinfo{journal}{Nat. Commun.} \textbf{\bibinfo{volume}{8}},
  \bibinfo{pages}{15897} (\bibinfo{year}{2017}).

\bibitem[{\citenamefont{Mitchell et~al.}(2010)\citenamefont{Mitchell, Richter,
  and Weidenm\"uller}}]{Mitchell_RMP_82_2845_2010}
\bibinfo{author}{\bibfnamefont{G.~E.} \bibnamefont{Mitchell}},
  \bibinfo{author}{\bibfnamefont{A.}~\bibnamefont{Richter}}, \bibnamefont{and}
  \bibinfo{author}{\bibfnamefont{H.~A.} \bibnamefont{Weidenm\"uller}},
  \bibinfo{journal}{Rev. Mod. Phys.} \textbf{\bibinfo{volume}{82}},
  \bibinfo{pages}{2845} (\bibinfo{year}{2010}).

\bibitem[{\citenamefont{Byrd et~al.}(2012)\citenamefont{Byrd, Harvey~Michels,
  Montgomery, C{\^o}t{\'e}, and Stwalley}}]{Byrd_JCP_136_014306_2012}
\bibinfo{author}{\bibfnamefont{J.~N.} \bibnamefont{Byrd}},
  \bibinfo{author}{\bibfnamefont{H.}~\bibnamefont{Harvey~Michels}},
  \bibinfo{author}{\bibfnamefont{J.~A.} \bibnamefont{Montgomery}},
  \bibinfo{author}{\bibfnamefont{R.}~\bibnamefont{C{\^o}t{\'e}}},
  \bibnamefont{and} \bibinfo{author}{\bibfnamefont{W.~C.}
  \bibnamefont{Stwalley}}, \bibinfo{journal}{J. Chem. Phys.}
  \textbf{\bibinfo{volume}{136}}, \bibinfo{pages}{014306}
  (\bibinfo{year}{2012}).

\bibitem[{\citenamefont{Idziaszek et~al.}(2010)\citenamefont{Idziaszek,
  Qu\'em\'ener, Bohn, and Julienne}}]{Idziaszek_PRA_82_020703_2010}
\bibinfo{author}{\bibfnamefont{Z.}~\bibnamefont{Idziaszek}},
  \bibinfo{author}{\bibfnamefont{G.}~\bibnamefont{Qu\'em\'ener}},
  \bibinfo{author}{\bibfnamefont{J.~L.} \bibnamefont{Bohn}}, \bibnamefont{and}
  \bibinfo{author}{\bibfnamefont{P.~S.} \bibnamefont{Julienne}},
  \bibinfo{journal}{Phys. Rev. A} \textbf{\bibinfo{volume}{82}},
  \bibinfo{pages}{020703} (\bibinfo{year}{2010}).

\end{thebibliography}

\onecolumngrid

\appendix

\section{Perturbative evaluation of the channels in the uncoupled representation}\label{app:a}

\subsection{For indistinguishable bosons}

Taking $m_{n_1}=m_{n_2}=m_l=m_{n_1}'=m_{n_2}'=m_l'=0$, and taking $l=0,2$,
Eq.~\eqref{Vdd} simplifies. We have
\begin{eqnarray*}
 \langle {n}_1 \ 0, {n}_2 \ 0, 0 \ 0 | V_\mathrm{dd} | {n}_1' \ 0, {n}_2' \ 0, 0 \ 0 \rangle
 =  0
\end{eqnarray*}
\begin{multline*}
 \langle {n}_1 \ 0, {n}_2 \ 0, 2 \ 0 | V_\mathrm{dd} | {n}_1' \ 0, {n}_2' \ 0, 2 \ 0 \rangle
 =  - \sqrt{30} \, \frac{d^2}{4 \pi \varepsilon_0 r^3}
  \, \left( \begin{array}{ccc}  1 & 1 & 2 \\ 0 & 0 & 0  \end{array}  \right) \\
\times  \sqrt{(2n_1+1) \, (2n_1'+1)}
\, \left( \begin{array}{ccc} n_1 & 1 & n_1' \\ 0 & 0 & 0 \end{array} \right)^2 \,
\,  \sqrt{(2n_2+1) \, (2n_2'+1)}
\, \left( \begin{array}{ccc} n_2 & 1 & n_2' \\ 0 & 0 & 0 \end{array} \right)^2
 \,  \sqrt{5} \sqrt{5}
\, \left( \begin{array}{ccc} 2 & 2 & 2 \\ 0 & 0 & 0 \end{array} \right)^2
\end{multline*}
and
\begin{multline*}
 \langle {n}_1 \ 0, {n}_2 \ 0, 0 \  0 | V_\mathrm{dd} | {n}_1' \ 0, {n}_2' \ 0, 2 \ 0 \rangle
 = \langle {n}_1 \ 0, {n}_2 \ 0, 2 \  0 | V_\mathrm{dd} | {n}_1' \ 0, {n}_2' \ 0, 0 \ 0 \rangle
 =  - \sqrt{30} \, \frac{d^2}{4 \pi \varepsilon_0 r^3}
  \, \left( \begin{array}{ccc}  1 & 1 & 2 \\ 0 & 0 & 0  \end{array}  \right) \\
\times  \sqrt{(2n_1+1) \, (2n_1'+1)}
\, \left( \begin{array}{ccc} n_1 & 1 & n_1' \\ 0 & 0 & 0 \end{array} \right)^2 \,
\,  \sqrt{(2n_2+1) \, (2n_2'+1)}
\, \left( \begin{array}{ccc} n_2 & 1 & n_2' \\ 0 & 0 & 0 \end{array} \right)^2
 \,  \sqrt{5}
\, \left( \begin{array}{ccc} 0 & 2 & 2 \\ 0 & 0 & 0 \end{array} \right)^2 .
\end{multline*}
Using Eq.~\eqref{ES1mol}, we find after some developments,
the expressions of the dipole-dipole interaction
between the intermediate channels $\big| \tilde{0} ,  \tilde{0} , l \big\rangle $ involved in the study
\begin{eqnarray*}
 \langle \tilde{0} , \tilde{0}, 0 | V_\mathrm{dd} | \tilde{0} , \tilde{0}, 0 \rangle
 =  0
\end{eqnarray*}
\begin{multline*}
 \langle \tilde{0} , \tilde{0}, 2 | V_\mathrm{dd} | \tilde{0} , \tilde{0}, 2 \rangle
 =  - \sqrt{30} \, \frac{d^2}{4 \pi \varepsilon_0 r^3}
 \, \left( \begin{array}{ccc}  1 & 1 & 2 \\ 0 & 0 & 0  \end{array}  \right) \times \\
\bigg\{ \,
 \alpha^2 \beta^2 \, \sqrt{3} \sqrt{3} \sqrt{5} \sqrt{5}
\, \left( \begin{array}{ccc} 0 & 1 & 1 \\ 0 & 0 & 0 \end{array} \right)^2 \,
\, \left( \begin{array}{ccc} 0 & 1 & 1 \\ 0 & 0 & 0 \end{array} \right)^2
\, \left( \begin{array}{ccc} 2 & 2 & 2 \\ 0 & 0 & 0 \end{array} \right)^2
+ \alpha^2 \beta^2 \, \sqrt{3} \sqrt{3} \sqrt{5} \sqrt{5}
\, \left( \begin{array}{ccc} 1 & 1 & 0 \\ 0 & 0 & 0 \end{array} \right)^2 \,
\, \left( \begin{array}{ccc} 1 & 1 & 0 \\ 0 & 0 & 0 \end{array} \right)^2
\, \left( \begin{array}{ccc} 2 & 2 & 2 \\ 0 & 0 & 0 \end{array} \right)^2 \\
+  \alpha^2 \beta^2 \, \sqrt{3} \sqrt{3} \sqrt{5} \sqrt{5}
\, \left( \begin{array}{ccc} 0 & 1 & 1 \\ 0 & 0 & 0 \end{array} \right)^2 \,
\, \left( \begin{array}{ccc} 1 & 1 & 0 \\ 0 & 0 & 0 \end{array} \right)^2
\, \left( \begin{array}{ccc} 2 & 2 & 2 \\ 0 & 0 & 0 \end{array} \right)^2
+  \alpha^2 \beta^2 \, \sqrt{3} \sqrt{3} \sqrt{5} \sqrt{5}
\, \left( \begin{array}{ccc} 1 & 1 & 0 \\ 0 & 0 & 0 \end{array} \right)^2 \,
\, \left( \begin{array}{ccc} 0 & 1 & 1 \\ 0 & 0 & 0 \end{array} \right)^2
\, \left( \begin{array}{ccc} 2 & 2 & 2 \\ 0 & 0 & 0 \end{array} \right)^2
\bigg\} \\
= - \frac{d^2}{4 \pi \varepsilon_0 r^3}
  \,  4 \, \alpha^2 \beta^2 \, (4/21)
\end{multline*}
and
\begin{multline*}
\langle \tilde{0} , \tilde{0}, 0 | V_\mathrm{dd} | \tilde{0} , \tilde{0}, 2 \rangle =
 \langle \tilde{0} , \tilde{0}, 2 | V_\mathrm{dd} | \tilde{0} , \tilde{0}, 0 \rangle
 =  - \sqrt{30} \, \frac{d^2}{4 \pi \varepsilon_0 r^3}
 \, \left( \begin{array}{ccc}  1 & 1 & 2 \\ 0 & 0 & 0  \end{array}  \right) \times \\
\bigg\{ \,
 \alpha^2 \beta^2 \, \sqrt{3} \sqrt{3} \sqrt{5}
\, \left( \begin{array}{ccc} 0 & 1 & 1 \\ 0 & 0 & 0 \end{array} \right)^2 \,
\, \left( \begin{array}{ccc} 0 & 1 & 1 \\ 0 & 0 & 0 \end{array} \right)^2
\, \left( \begin{array}{ccc} 0 & 2 & 2 \\ 0 & 0 & 0 \end{array} \right)^2
+ \alpha^2 \beta^2 \, \sqrt{3} \sqrt{3} \sqrt{5}
\, \left( \begin{array}{ccc} 1 & 1 & 0 \\ 0 & 0 & 0 \end{array} \right)^2 \,
\, \left( \begin{array}{ccc} 1 & 1 & 0 \\ 0 & 0 & 0 \end{array} \right)^2
\, \left( \begin{array}{ccc} 0 & 2 & 2 \\ 0 & 0 & 0 \end{array} \right)^2 \\
+  \alpha^2 \beta^2 \, \sqrt{3} \sqrt{3} \sqrt{5}
\, \left( \begin{array}{ccc} 0 & 1 & 1 \\ 0 & 0 & 0 \end{array} \right)^2 \,
\, \left( \begin{array}{ccc} 1 & 1 & 0 \\ 0 & 0 & 0 \end{array} \right)^2
\, \left( \begin{array}{ccc} 0 & 2 & 2 \\ 0 & 0 & 0 \end{array} \right)^2
+  \alpha^2 \beta^2 \, \sqrt{3} \sqrt{3} \sqrt{5}
\, \left( \begin{array}{ccc} 1 & 1 & 0 \\ 0 & 0 & 0 \end{array} \right)^2 \,
\, \left( \begin{array}{ccc} 0 & 1 & 1 \\ 0 & 0 & 0 \end{array} \right)^2
\, \left( \begin{array}{ccc} 0 & 2 & 2 \\ 0 & 0 & 0 \end{array} \right)^2
\bigg\} \\
= - \frac{d^2}{4 \pi \varepsilon_0 r^3}
  \,   4 \, \alpha^2 \beta^2 \, (2/ \, 3\sqrt{5}) .
\end{multline*}
We end up with a two-by-two matrix
\begin{eqnarray*}
\left[ \begin{array}{cc}  {E}_1 &
{\cal W}   \\
{\cal W}  &
{E}_2
\end{array}\right] =
\left[ \begin{array}{cc}
 2 E_{\tilde{0}} - \frac{C_6^\mathrm{el}}{r^6}  &
- \frac{d^2}{4 \pi \varepsilon_0 r^3}
 \,  4 \, \alpha^2 \beta^2 \, (2/ \, 3\sqrt{5})   \\
- \frac{d^2}{4 \pi \varepsilon_0 r^3}
  \,  4 \, \alpha^2 \beta^2 \, (2/ \, 3\sqrt{5})  &
2 E_{\tilde{0}} - \frac{C_6^\mathrm{el}}{r^6} + \frac{6 \hbar^2}{2  \mu  r^2} - \frac{d^2}{4 \pi \varepsilon_0 r^3}    4 \, \alpha^2 \beta^2 \, (4/21)
\end{array}\right]
\end{eqnarray*}
which can be easily diagonalized and lead to the corresponding lowest eigenstate,
noted $\big| \Omega({\cal E}) \big\rangle = \big| \tilde{n}_1 ,  \tilde{n}_2, \tilde{l} \big\rangle$
and expressed as a function of $\big| \tilde{n}_1 , \tilde{n}_2, l \big\rangle$.
For a given ${\cal E}$, if ${E}_2 \ge {E}_1$, it is given by
\begin{eqnarray*}
\big| \Omega({\cal E}) \big\rangle = \big| \tilde{0} ,  \tilde{0}, \tilde{0} \big\rangle =
\cos(\eta{/2}) \, \big| \tilde{0} ,  \tilde{0}, {0} \big\rangle
- \sin(\eta{/2})   \, \big| \tilde{0} ,  \tilde{0}, 2 \big\rangle
\end{eqnarray*}
and if ${E}_2 < {E}_1$, it is given by
\begin{eqnarray*}
\big| \Omega({\cal E}) \big\rangle = \big| \tilde{0} ,  \tilde{0}, \tilde{0} \big\rangle =
- \sin(\eta{/2})   \, \big| \tilde{0} ,  \tilde{0}, {0} \big\rangle
+ \cos(\eta{/2})   \, \big| \tilde{0} ,  \tilde{0}, 2 \big\rangle
\end{eqnarray*}
with $\eta$ defined in Eq.~\eqref{eta-B}.
From the expression of $\big| \tilde{n}_1 , \tilde{n}_2, l \big\rangle$
as a function of $\big| {n}_1, {n}_2, l \big\rangle$
using Eq.~\eqref{ES1mol}, we have
\begin{eqnarray*}
\big| \tilde{0} ,  \tilde{0}, {0} \big\rangle  &=&
\cos^2(\theta{/2}) \, \big| {0} , {0}, {0} \big\rangle
- \frac{1}{2} \, \sin\theta \, \big| {0} , {1}, {0} \big\rangle
- \frac{1}{2} \, \sin\theta \, \big| {1} , {0}, {0} \big\rangle
+ \sin^2(\theta{/2}) \, \big| {1} , {1}, {0} \big\rangle  \nonumber \\
\big| \tilde{0} ,  \tilde{0}, {2} \big\rangle  &=&
\cos^2(\theta{/2}) \, \big| {0} , {0}, {2} \big\rangle
- \frac{1}{2} \, \sin\theta \, \big| {0} , {1}, {2} \big\rangle
- \frac{1}{2} \, \sin\theta \, \big| {1} , {0}, {2} \big\rangle
+  \sin^2(\theta{/2}) \,  \big| {1} , {1}, {2} \big\rangle .
\end{eqnarray*}
The kets $\big| {n_1} , {n_2}, {l} \big\rangle $
as a function of the kets $\big| (n_1, n_2) \, n_{12}  \, l \, ; J \, M \big\rangle $
are given in Eq.~\eqref{n1n2l-B} below.

\subsection{For indistinguishable fermions}

We take $m_{n_1}=m_{n_2}=m_l=m_{n_1}'=m_{n_2}'=m_l'=0$, and $l=1,3$.
We only consider the attractive, head-to-tail approach $m_l=0$ of the $l=1$ p-wave collision.
The side-by-side approach $m_l=\pm1$ will give a repulsive interaction
as the electric field is increased and its contribution to the dynamics can be
ignored as a good approximation~\cite{Quemener_PRA_84_062703_2011}.
Eq.~\eqref{Vdd} simplifies and we have
\begin{multline*}
 \langle {n}_1 \ 0, {n}_2 \ 0, 1 \  0 | V_\mathrm{dd} | {n}_1'\ 0, {n}_2' \ 0, 1 \ 0 \rangle
 =  - \sqrt{30} \, \frac{d^2}{4 \pi \varepsilon_0 r^3}
  \, \left( \begin{array}{ccc}  1 & 1 & 2 \\ 0 & 0 & 0  \end{array}  \right) \\
\times  \sqrt{(2n_1+1) \, (2n_1'+1)}
\, \left( \begin{array}{ccc} n_1 & 1 & n_1' \\ 0 & 0 & 0 \end{array} \right)^2 \,
\,  \sqrt{(2n_2+1) \, (2n_2'+1)}
\, \left( \begin{array}{ccc} n_2 & 1 & n_2' \\ 0 & 0 & 0 \end{array} \right)^2
 \,  \sqrt{3} \sqrt{3}
\, \left( \begin{array}{ccc} 1 & 2 & 1 \\ 0 & 0 & 0 \end{array} \right)^2
\end{multline*}
\begin{multline*}
 \langle {n}_1 \ 0, {n}_2 \ 0, 3 \  0 | V_\mathrm{dd} | {n}_1' \ 0, {n}_2' \ 0, 3 \ 0 \rangle
 =  - \sqrt{30} \, \frac{d^2}{4 \pi \varepsilon_0 r^3}
  \, \left( \begin{array}{ccc}  1 & 1 & 2 \\ 0 & 0 & 0  \end{array}  \right) \\
\times  \sqrt{(2n_1+1) \, (2n_1'+1)}
\, \left( \begin{array}{ccc} n_1 & 1 & n_1' \\ 0 & 0 & 0 \end{array} \right)^2 \,
\,  \sqrt{(2n_2+1) \, (2n_2'+1)}
\, \left( \begin{array}{ccc} n_2 & 1 & n_2' \\ 0 & 0 & 0 \end{array} \right)^2
 \,  \sqrt{7} \sqrt{7}
\, \left( \begin{array}{ccc} 3 & 2 & 3 \\ 0 & 0 & 0 \end{array} \right)^2
\end{multline*}
and
\begin{multline*}
 \langle {n}_1 \ 0, {n}_2 \ 0, 1 \ 0 | V_\mathrm{dd} | {n}_1' \ 0, {n}_2' \ 0, 3 \ 0 \rangle
 = \langle {n}_1 \ 0, {n}_2 \ 0, 3 \ 0 | V_\mathrm{dd} | {n}_1' \ 0, {n}_2' \ 0, 1 \ 0 \rangle
 =  - \sqrt{30} \, \frac{d^2}{4 \pi \varepsilon_0 r^3}
  \, \left( \begin{array}{ccc}  1 & 1 & 2 \\ 0 & 0 & 0  \end{array}  \right) \\
\times  \sqrt{(2n_1+1) \, (2n_1'+1)}
\, \left( \begin{array}{ccc} n_1 & 1 & n_1' \\ 0 & 0 & 0 \end{array} \right)^2 \,
\,  \sqrt{(2n_2+1) \, (2n_2'+1)}
\, \left( \begin{array}{ccc} n_2 & 1 & n_2' \\ 0 & 0 & 0 \end{array} \right)^2
 \,  \sqrt{3}  \sqrt{7}
\, \left( \begin{array}{ccc} 1 & 2 & 3 \\ 0 & 0 & 0 \end{array} \right)^2 .
\end{multline*}
Using Eq.~\eqref{ES1mol}, we find then for the intermediate channels
$\big| \tilde{0} ,  \tilde{0} , l \big\rangle $
\begin{multline*}
 \langle \tilde{0} , \tilde{0}, 1 | V_\mathrm{dd} | \tilde{0} , \tilde{0}, 1 \rangle
 =  - \sqrt{30} \, \frac{d^2}{4 \pi \varepsilon_0 r^3}
 \, \left( \begin{array}{ccc}  1 & 1 & 2 \\ 0 & 0 & 0  \end{array}  \right) \times \\
\bigg\{ \,
 \alpha^2 \beta^2 \, \sqrt{3} \sqrt{3} \sqrt{3} \sqrt{3}
\, \left( \begin{array}{ccc} 0 & 1 & 1 \\ 0 & 0 & 0 \end{array} \right)^2 \,
\, \left( \begin{array}{ccc} 0 & 1 & 1 \\ 0 & 0 & 0 \end{array} \right)^2
\, \left( \begin{array}{ccc} 1 & 2 & 1 \\ 0 & 0 & 0 \end{array} \right)^2
+ \alpha^2 \beta^2 \, \sqrt{3} \sqrt{3} \sqrt{3} \sqrt{3}
\, \left( \begin{array}{ccc} 1 & 1 & 0 \\ 0 & 0 & 0 \end{array} \right)^2 \,
\, \left( \begin{array}{ccc} 1 & 1 & 0 \\ 0 & 0 & 0 \end{array} \right)^2
\, \left( \begin{array}{ccc} 1 & 2 & 1 \\ 0 & 0 & 0 \end{array} \right)^2 \\
+  \alpha^2 \beta^2 \, \sqrt{3} \sqrt{3} \sqrt{3} \sqrt{3}
\, \left( \begin{array}{ccc} 0 & 1 & 1 \\ 0 & 0 & 0 \end{array} \right)^2 \,
\, \left( \begin{array}{ccc} 1 & 1 & 0 \\ 0 & 0 & 0 \end{array} \right)^2
\, \left( \begin{array}{ccc} 1 & 2 & 1 \\ 0 & 0 & 0 \end{array} \right)^2
+  \alpha^2 \beta^2 \, \sqrt{3} \sqrt{3} \sqrt{3} \sqrt{3}
\, \left( \begin{array}{ccc} 1 & 1 & 0 \\ 0 & 0 & 0 \end{array} \right)^2 \,
\, \left( \begin{array}{ccc} 0 & 1 & 1 \\ 0 & 0 & 0 \end{array} \right)^2
\, \left( \begin{array}{ccc} 1 & 2 & 1 \\ 0 & 0 & 0 \end{array} \right)^2
\bigg\} \\
= - \frac{d^2}{4 \pi \varepsilon_0 r^3}
 \,  4 \, \alpha^2 \beta^2 \, (4/15)
\end{multline*}
\begin{multline*}
 \langle \tilde{0} , \tilde{0}, 3 | V_\mathrm{dd} | \tilde{0} , \tilde{0}, 3 \rangle
 =  - \sqrt{30} \, \frac{d^2}{4 \pi \varepsilon_0 r^3}
 \, \left( \begin{array}{ccc}  1 & 1 & 2 \\ 0 & 0 & 0  \end{array}  \right) \times \\
\bigg\{ \,
 \alpha^2 \beta^2 \, \sqrt{3} \sqrt{3} \sqrt{7} \sqrt{7}
\, \left( \begin{array}{ccc} 0 & 1 & 1 \\ 0 & 0 & 0 \end{array} \right)^2 \,
\, \left( \begin{array}{ccc} 0 & 1 & 1 \\ 0 & 0 & 0 \end{array} \right)^2
\, \left( \begin{array}{ccc} 3 & 2 & 3 \\ 0 & 0 & 0 \end{array} \right)^2
+ \alpha^2 \beta^2 \, \sqrt{3} \sqrt{3} \sqrt{7} \sqrt{7}
\, \left( \begin{array}{ccc} 1 & 1 & 0 \\ 0 & 0 & 0 \end{array} \right)^2 \,
\, \left( \begin{array}{ccc} 1 & 1 & 0 \\ 0 & 0 & 0 \end{array} \right)^2
\, \left( \begin{array}{ccc} 3 & 2 & 3 \\ 0 & 0 & 0 \end{array} \right)^2 \\
+  \alpha^2 \beta^2 \, \sqrt{3} \sqrt{3} \sqrt{7} \sqrt{7}
\, \left( \begin{array}{ccc} 0 & 1 & 1 \\ 0 & 0 & 0 \end{array} \right)^2 \,
\, \left( \begin{array}{ccc} 1 & 1 & 0 \\ 0 & 0 & 0 \end{array} \right)^2
\, \left( \begin{array}{ccc} 3 & 2 & 3 \\ 0 & 0 & 0 \end{array} \right)^2
+  \alpha^2 \beta^2 \, \sqrt{3} \sqrt{3} \sqrt{7} \sqrt{7}
\, \left( \begin{array}{ccc} 1 & 1 & 0 \\ 0 & 0 & 0 \end{array} \right)^2 \,
\, \left( \begin{array}{ccc} 0 & 1 & 1 \\ 0 & 0 & 0 \end{array} \right)^2
\, \left( \begin{array}{ccc} 3 & 2 & 3 \\ 0 & 0 & 0 \end{array} \right)^2
\bigg\} \\
= - \frac{d^2}{4 \pi \varepsilon_0 r^3}
 \,   4 \, \alpha^2 \beta^2 \, (8/45)
\end{multline*}
and
\begin{multline*}
  \langle \tilde{0} , \tilde{0}, 1 | V_\mathrm{dd} | \tilde{0} , \tilde{0}, 3 \rangle =
 \langle \tilde{0} , \tilde{0}, 3 | V_\mathrm{dd} | \tilde{0} , \tilde{0}, 1 \rangle
 =  - \sqrt{30} \, \frac{d^2}{4 \pi \varepsilon_0 r^3}
 \, \left( \begin{array}{ccc}  1 & 1 & 2 \\ 0 & 0 & 0  \end{array}  \right) \times \\
\bigg\{ \,
 \alpha^2 \beta^2 \, \sqrt{3} \sqrt{3} \sqrt{3} \sqrt{7}
\, \left( \begin{array}{ccc} 0 & 1 & 1 \\ 0 & 0 & 0 \end{array} \right)^2 \,
\, \left( \begin{array}{ccc} 0 & 1 & 1 \\ 0 & 0 & 0 \end{array} \right)^2
\, \left( \begin{array}{ccc} 1 & 2 & 3 \\ 0 & 0 & 0 \end{array} \right)^2
+ \alpha^2 \beta^2 \, \sqrt{3} \sqrt{3} \sqrt{3} \sqrt{7}
\, \left( \begin{array}{ccc} 1 & 1 & 0 \\ 0 & 0 & 0 \end{array} \right)^2 \,
\, \left( \begin{array}{ccc} 1 & 1 & 0 \\ 0 & 0 & 0 \end{array} \right)^2
\, \left( \begin{array}{ccc} 1 & 2 & 3 \\ 0 & 0 & 0 \end{array} \right)^2 \\
+  \alpha^2 \beta^2 \, \sqrt{3} \sqrt{3} \sqrt{3} \sqrt{7}
\, \left( \begin{array}{ccc} 0 & 1 & 1 \\ 0 & 0 & 0 \end{array} \right)^2 \,
\, \left( \begin{array}{ccc} 1 & 1 & 0 \\ 0 & 0 & 0 \end{array} \right)^2
\, \left( \begin{array}{ccc} 1 & 2 & 3 \\ 0 & 0 & 0 \end{array} \right)^2
+  \alpha^2 \beta^2 \, \sqrt{3} \sqrt{3} \sqrt{3} \sqrt{7}
\, \left( \begin{array}{ccc} 1 & 1 & 0 \\ 0 & 0 & 0 \end{array} \right)^2 \,
\, \left( \begin{array}{ccc} 0 & 1 & 1 \\ 0 & 0 & 0 \end{array} \right)^2
\, \left( \begin{array}{ccc} 1 & 2 & 3 \\ 0 & 0 & 0 \end{array} \right)^2
\bigg\} \\
= - \frac{d^2}{4 \pi \varepsilon_0 r^3}
  \,   4 \, \alpha^2 \beta^2 \, (2 \sqrt{3} \, / \, 5\sqrt{7}) .
\end{multline*}
We end up with a two-by-two matrix
\begin{eqnarray*}
\left[ \begin{array}{cc}  {E}_1 &
{\cal W}   \\
{\cal W}  &
{E}_2
\end{array}\right] =
\left[ \begin{array}{cc}
2 E_{\tilde{0}} - \frac{C_6^\mathrm{el}}{r^6} + \frac{2 \hbar^2}{2  \mu  r^2} - \frac{d^2}{4 \pi \varepsilon_0 r^3} \, 4 \, \alpha^2 \beta^2 \, (4/15) &
- \frac{d^2}{4 \pi \varepsilon_0 r^3}
 \,  4 \, \alpha^2 \beta^2 \, (2 \sqrt{3} \, / \, 5\sqrt{7})   \\
- \frac{d^2}{4 \pi \varepsilon_0 r^3}
 \,   4 \, \alpha^2 \beta^2 \, (2 \sqrt{3} \, / \, 5\sqrt{7})  &
2 E_{\tilde{0}} - \frac{C_6^\mathrm{el}}{r^6} + \frac{12 \hbar^2}{2  \mu  r^2} - \frac{d^2}{4 \pi \varepsilon_0 r^3} \, 4 \, \alpha^2 \beta^2 \, (8/45)
\end{array}\right]
\end{eqnarray*}
which can be easily diagonalized and lead to the corresponding lowest eigenstate
noted $\big| \Omega({\cal E}) \big\rangle = \big| \tilde{n}_1 ,  \tilde{n}_2, \tilde{l} \big\rangle$
and expressed as a function of $\big| \tilde{n}_1 , \tilde{n}_2, l \big\rangle$.
For a given ${\cal E}$, if ${E}_2 \ge {E}_1$, it is given by
\begin{eqnarray*}
\big| \Omega({\cal E}) \big\rangle = \big| \tilde{0} ,  \tilde{0}, \tilde{1} \big\rangle =
\cos(\eta{/2}) \, \big| \tilde{0} ,  \tilde{0}, {1} \big\rangle
- \sin(\eta{/2})   \, \big| \tilde{0} ,  \tilde{0}, 3 \big\rangle
\end{eqnarray*}
and if ${E}_2 < {E}_1$, it is given by
\begin{eqnarray*}
\big| \Omega({\cal E}) \big\rangle = \big| \tilde{0} ,  \tilde{0}, \tilde{1} \big\rangle =
- \sin(\eta{/2})   \, \big| \tilde{0} ,  \tilde{0}, {1} \big\rangle
+ \cos(\eta{/2})   \, \big| \tilde{0} ,  \tilde{0}, 3 \big\rangle
\end{eqnarray*}
with $\eta$ defined in Eq.~\eqref{eta-F}.
From the expression of $\big| \tilde{n}_1 , \tilde{n}_2, l \big\rangle$
as a function of $\big| {n}_1, {n}_2, l \big\rangle$
using Eq.~\eqref{ES1mol}, we have
\begin{eqnarray*}
\big| \tilde{0} ,  \tilde{0}, {1} \big\rangle &=&
\cos^2(\theta{/2}) \, \big| {0} , {0}, {1} \big\rangle
- \frac{1}{2} \, \sin\theta \, \big| {0} , {1}, {1} \big\rangle
- \frac{1}{2} \, \sin\theta \, \big| {1} , {0}, {1} \big\rangle
+ \sin^2(\theta{/2}) \, \big| {1} , {1}, {1} \big\rangle  \nonumber \\
\big| \tilde{0} ,  \tilde{0}, {3} \big\rangle &=&
\cos^2(\theta{/2}) \, \big| {0} , {0}, {3} \big\rangle
- \frac{1}{2} \, \sin\theta \, \big| {0} , {1}, {3} \big\rangle
- \frac{1}{2} \, \sin\theta \, \big| {1} , {0}, {3} \big\rangle
+  \sin^2(\theta{/2}) \,  \big| {1} , {1}, {3} \big\rangle .
\end{eqnarray*}
The kets $\big| {n_1} , {n_2}, {l} \big\rangle $
as a function of the kets $\big| (n_1, n_2) \, n_{12}  \, l \, ; J \, M \big\rangle $
are given in Eq.~\eqref{n1n2l-F} below.

\section{Link between coupled and uncoupled representation}\label{app:b}

We express the relevant kets $\big| n_1  ,  n_2  \big\rangle$
as a function of the kets $\big| (n_1, n_2 ) \, n_{12} \big\rangle $
using the usual transformation
\begin{eqnarray}
\big| n_1  \, m_{n_1},  n_2  \, m_{n_2} \big\rangle = \sum_{n_{12}} \sum_{m_{n_{12}}}
\big| n_1 \, n_2 \, ; n_{12}  \, m_{n_{12}} \big\rangle 
\langle n_1 \, n_2 \,  ; n_{12}  \, m_{n_{12}} \big| n_1  \, m_{n_1},  n_2 \, m_{n_2} \big\rangle .
\label{COUPL1}
\end{eqnarray}
We omit the notation of the projection numbers as they are all zero in this study.
We find
\begin{align*}
\big| 0  ,  0  \big\rangle &= \big| (0, 0) \,  0  \big\rangle
& \big| 1  ,  1  \big\rangle &= - \sqrt{\frac{1}{3}} \, \big| (1, 1) \, 0  \big\rangle + \sqrt{\frac{2}{3}} \, \big| (1, 1 ) \, 2  \big\rangle  \\
\big| 0  ,  1  \big\rangle &= \big| (0, 1 ) \, 1  \big\rangle
& \big| 1  ,  0  \big\rangle &= \big| (1, 0) \, 1  \big\rangle .
\end{align*}
We then express the kets $\big| (n_1, n_2) \, n_{12},  l \big\rangle$
as a function of the kets $\big| (n_1, n_2) \, n_{12}  \, l \, ; J \, M \big\rangle $
using 
\begin{eqnarray}
\big| (n_1, n_2) \, n_{12}  \, m_{n_{12}},  l \, m_l \big\rangle = \sum_{J} \sum_{M_J}
\big| (n_1, n_2) \, n_{12}  \, l \, ; J \, M \big\rangle  
\langle (n_1, n_2) \, n_{12}  \, l \,  ; J \, M \big| (n_1, n_2) \, n_{12}  \, m_{n_{12}},  l \, m_l \big\rangle .
\label{COUPL2}
\end{eqnarray}
We find
\begin{align*}
\big| (0, 0) \, 0  ,  0  \big\rangle &=  \big| (0, 0) \, 0  \, 0 \, ; 0 \, 0 \big\rangle
&  \big| (1, 1) \, 0  ,  0  \big\rangle &=  \big| (1, 1) \, 0  \, 0 \, ; 0 \, 0 \big\rangle  \\
\big| (0, 0) \, 0  ,  2  \big\rangle &=  \big| (0, 0) \, 0  \, 2 \, ; 2 \, 0 \big\rangle
& \big| (1, 1) \, 0  ,  2  \big\rangle &=  \big| (1, 1) \, 0  \, 2 \, ; 2 \, 0 \big\rangle  \\
\big| (0, 0) \, 0  ,  1  \big\rangle &=  \big| (0, 0) \, 0  \, 1 \, ; 1 \, 0 \big\rangle
& \big| (1, 1) \, 0  ,  1  \big\rangle &=  \big| (1, 1) \, 0  \, 1 \, ; 1 \, 0 \big\rangle  \\
\big| (0, 0) \, 0  ,  3  \big\rangle &=  \big| (0, 0) \, 0  \, 3 \, ; 3 \, 0 \big\rangle
& \big| (1, 1) \, 0  ,  3  \big\rangle &=  \big| (1, 1) \, 0  \, 3 \, ; 3 \, 0 \big\rangle
\end{align*}
\begin{align*}
\big| (0, 1) \, 1  ,  0  \big\rangle &=  \big| (0, 1) \, 1  \, 0 \, ; 1 \, 0 \big\rangle
&  \big| (1, 0) \, 1  ,  0  \big\rangle &=  \big| (1, 0) \, 1  \, 0 \, ; 1 \, 0 \big\rangle  \\
\big| (0, 1) \, 1  ,  2  \big\rangle &= - \sqrt{\frac{2}{5}}  \, \big| (0, 1) \, 1  \, 2 \, ; 1 \, 0 \big\rangle
+  \sqrt{\frac{3}{5}} \, \big| (0, 1) \, 1  \, 2 \, ; 3 \, 0 \big\rangle
& \big| (1, 0) \, 1  ,  2  \big\rangle &= - \sqrt{\frac{2}{5}}  \, \big| (1, 0) \, 1  \, 2 \, ; 1 \, 0 \big\rangle
+  \sqrt{\frac{3}{5}} \, \big| (1, 0) \, 1  \, 2 \, ; 3 \, 0 \big\rangle  \\
\big| (0, 1) \, 1  ,  1  \big\rangle &= - \sqrt{\frac{1}{3}}  \, \big| (0, 1) \, 1  \, 1 \, ; 0 \, 0 \big\rangle
+  \sqrt{\frac{2}{3}}  \, \big| (0, 1) \, 1  \, 1 \, ; 2 \, 0 \big\rangle
&  \big| (1, 0) \, 1  ,  1  \big\rangle &= - \sqrt{\frac{1}{3}}  \, \big| (1, 0) \, 1  \, 1 \, ; 0 \, 0 \big\rangle
+  \sqrt{\frac{2}{3}}  \, \big| (1, 0) \, 1  \, 1 \, ; 2 \, 0 \big\rangle \\
\big| (0, 1) \, 1  ,  3  \big\rangle &=  - \sqrt{\frac{3}{7}}  \, \big| (0, 1) \, 1  \, 3 \, ; 2 \, 0 \big\rangle
+  \sqrt{\frac{4}{7}}  \, \big| (0, 1) \, 1  \, 3 \, ; 4 \, 0 \big\rangle
& \big| (1, 0) \, 1  ,  3  \big\rangle &=  - \sqrt{\frac{3}{7}}  \, \big| (1, 0) \, 1  \, 3 \, ; 2 \, 0 \big\rangle
+  \sqrt{\frac{4}{7}} \, \big| (1, 0) \, 1  \, 3 \, ; 4 \, 0 \big\rangle
\end{align*}
\begin{eqnarray*}
\big| (1, 1) \, 2  ,  0  \big\rangle &=&  \big| (1, 1) \, 2  \, 0 \, ; 2 \, 0 \big\rangle  \nonumber \\
 \big| (1, 1) \, 2  ,  2  \big\rangle &=& \sqrt{\frac{1}{5}}  \, \big| (1, 1) \, 2  \, 2 \, ; 0 \, 0 \big\rangle
- \sqrt{\frac{2}{7}} \, \big| (1, 1) \, 2  \, 2 \, ; 2 \, 0 \big\rangle
+ 3 \, \sqrt{\frac{2}{35}}  \, \big| (1, 1) \, 2  \, 2 \, ; 4 \, 0 \big\rangle   \nonumber \\
\big| (1, 1) \, 2  ,  1  \big\rangle &=& { - } \sqrt{\frac{2}{5}}  \, \big| (1, 1) \, 2  \, 1 \, ; 1 \, 0 \big\rangle
+  \sqrt{\frac{3}{5}}  \, \big| (1, 1) \, 2  \, 1 \, ; 3 \, 0 \big\rangle \nonumber \\
\big| (1, 1) \, 2  ,  3  \big\rangle &=& \sqrt{\frac{9}{35}}  \, \big| (1, 1) \, 2  \, 3 \, ; 1 \, 0 \big\rangle
- \sqrt{\frac{4}{15}}\, \big| (1, 1) \, 2  \, 3 \, ; 3 \, 0 \big\rangle
+ \sqrt{\frac{10}{21}}  \, \big| (1, 1) \, 2  \, 3 \, ; 5 \, 0 \big\rangle .
\end{eqnarray*}
\\

Using the above equations, we find the kets $\big| {n_1} , {n_2}, l \big\rangle$
as a function of the kets $\big| (n_1, n_2) \, n_{12}  \, l \, ; J \, M \big\rangle $
for indistinguishable bosons
\begin{eqnarray}
\big| {0} , {0}, {0} \big\rangle &=&  \big| (0, 0) \, 0  \, 0 \, ; 0 \, 0 \big\rangle \nonumber \\
\big| {0} , {1}, {0} \big\rangle &=&  \big| (0, 1) \, 1  \, 0 \, ; 1 \, 0 \big\rangle \nonumber \\
\big| {1} , {0}, {0} \big\rangle &=&  \big| (1, 0) \, 1  \, 0 \, ; 1 \, 0 \big\rangle \nonumber \\
\big| {1} , {1}, {0} \big\rangle &=&
- \sqrt{1/3} \, \big| (1, 1) \, 0  \, 0 \, ; 0 \, 0 \big\rangle
 { + } \sqrt{2/3} \, \big| (1, 1) \, 2  \, 0 \, ; 2 \, 0 \big\rangle  \nonumber \\
\big| {0} , {0}, {2} \big\rangle &=&  \big| (0, 0) \, 0  \, 2 \, ; 2 \, 0 \big\rangle \nonumber \\
\big| {0} , {1}, {2} \big\rangle &=&
- \sqrt{2/5} \, \big| (0, 1) \, 1  \, 2 \, ; 1 \, 0 \big\rangle
+ \sqrt{3/5} \times \big| (0, 1) \, 1  \, 2 \, ; 3 \, 0 \big\rangle \nonumber \\
\big| {1} , {0}, {2} \big\rangle &=&
- \sqrt{2/5} \, \big| (1, 0) \, 1  \, 2 \, ; 1 \, 0 \big\rangle
+ \sqrt{3/5} \times \big| (1, 0) \, 1  \, 2 \, ; 3 \, 0 \big\rangle \nonumber \\
\big| {1} , {1}, {2} \big\rangle
&=& - \sqrt{1/3} \, \big| (1, 1) \, 0  \, 2 \, ; 2 \, 0 \big\rangle
{ + } \sqrt{2/15} \, \big| (1, 1) \, 2  \, 2 \, ; 0 \, 0 \big\rangle { - }  \sqrt{4/21}  \, \big| (1, 1) \, 2  \, 2 \, ; 2 \, 0 \big\rangle  { + }  \sqrt{12/35} \, \big| (1, 1) \, 2  \, 2 \, ; 4 \, 0 \big\rangle   .
\label{n1n2l-B}
\end{eqnarray}
\\

Similarly for indistinguishable fermions, we find
\begin{eqnarray}
\big| {0} , {0}, {1} \big\rangle &=&  \big| (0, 0) \, 0  \, 1 \, ; 1 \, 0 \big\rangle \nonumber \\
\big| {0} , {1}, {1} \big\rangle &=&
- \sqrt{1/3} \, \big| (0, 1) \, 1  \, 1 \, ; 0 \, 0 \big\rangle + \sqrt{2/3} \,  \big| (0, 1) \, 1  \, 1 \, ; 2 \, 0 \big\rangle \nonumber \\
\big| {1} , {0}, {1} \big\rangle &=&
- \sqrt{1/3} \, \big| (1, 0) \, 1  \, 1 \, ; 0 \, 0 \big\rangle + \sqrt{2/3} \,  \big| (1, 0) \, 1  \, 1 \, ; 2 \, 0 \big\rangle \nonumber \\
\big| {1} , {1}, {1} \big\rangle &=&
 - \sqrt{1/3} \, \big| (1, 1) \, 0  \, 1 \, ; 1 \, 0 \big\rangle
{ - } \sqrt{4/15} \, \big| (1, 1) \, 2  \, 1 \, ; 1 \, 0 \big\rangle
{ + } \sqrt{2/5} \, \big| (1, 1) \, 2  \, 1 \, ; 3 \, 0 \big\rangle  \nonumber \\
\big| {0} , {0}, {3} \big\rangle &=&  \big| (0, 0) \, 0  \, 3 \, ; 3 \, 0 \big\rangle \nonumber \\
\big| {0} , {1}, {3} \big\rangle &=&
- \sqrt{3/7} \, \big| (0, 1) \, 1  \, 3 \, ; 2 \, 0 \big\rangle
+ \sqrt{4/7} \times \big| (0, 1) \, 1  \, 3 \, ; 4 \, 0 \big\rangle \nonumber \\
\big| {1} , {0}, {3} \big\rangle &=&
- \sqrt{3/7} \, \big| (1, 0) \, 1  \, 3 \, ; 2 \, 0 \big\rangle
+ \sqrt{4/7} \times \big| (1, 0) \, 1  \, 3 \, ; 4 \, 0 \big\rangle \nonumber \\
\big| {1} , {1}, {3} \big\rangle &=&
- \sqrt{1/3} \, \big| (1, 1) \, 0  \, 3 \, ; 3 \, 0 \big\rangle
+ { \sqrt{6/35} } \, \big| (1, 1) \, 2  \, 3 \, ; 1 \, 0 \big\rangle
{ - } \sqrt{8/45}  \, \big| (1, 1) \, 2  \, 3 \, ; 3 \, 0 \big\rangle
{ + } \sqrt{20/63} \, \big| (1, 1) \, 2  \, 3 \, ; 5 \, 0 \big\rangle      .
\label{n1n2l-F}
\end{eqnarray}

\section{Evaluation of the probabilities}\label{app:c}

Projecting the kets $\big| (n_1, n_2) \, n_{12}  \, l \, ; J \, M \big\rangle $
onto the dressed eigenstates $\big| \tilde{n}_1 , \tilde{n}_2,  \tilde{l} \big\rangle$
and taking the modulus squared, we get the corresponding probabilities
to find the admixture of a $J$ component due to the dressing by the electric field
seen by the individual molecules.
For each $J$, one can sum these contributions to get the $J$-dependent probability $P_J^B({\cal E})$ for bosons
or $P_J^F({\cal E})$ for fermions.
This is plotted in Fig.~\ref{FIG-PROBA-ADIM} for different $J$ components.

\subsection{For indistinguishable bosons}

For bosons, when ${\cal E}$ and $\tilde{r}$ are such that ${E}_2 \ge {E}_1$, the probabilities are  \\

\noindent for $J=0, M=0$
\begin{eqnarray}
\big| \big\langle  (0, 0) \, 0  \, 0 \, ; 0 \, 0 \big| \tilde{0} , \tilde{0},  \tilde{0} \big\rangle \big|^2
&=& \cos^2(\eta{/2}) \, \cos^4(\theta{/2}) \\
\big| \big\langle  (1, 1) \, 0  \, 0 \, ; 0 \, 0 \big| \tilde{0} , \tilde{0},  \tilde{0} \big\rangle \big|^2
&=& \cos^2(\eta{/2}) \, \sin^4(\theta{/2}) \times (1/3)   \nonumber \\
\big| \big\langle  (1, 1) \, 2  \, 2 \, ; 0 \, 0 \big| \tilde{0} , \tilde{0},  \tilde{0} \big\rangle \big|^2
&=& \sin^2(\eta{/2}) \, \sin^4(\theta{/2}) \times (2/15)  \nonumber  ,
\label{P0-B}
\end{eqnarray}
for $J=1, M=0$
\begin{eqnarray*}
\big| \big\langle  (0, 1) \, 1  \, 0 \, ; 1 \, 0 \big| \tilde{0} , \tilde{0},  \tilde{0} \big\rangle \big|^2
&=& \cos^2(\eta{/2}) \, (\sin^2\theta)/4 \nonumber \\
\big| \big\langle  (1, 0) \, 1  \, 0 \, ; 1 \, 0 \big| \tilde{0} , \tilde{0},  \tilde{0} \big\rangle \big|^2
&=& \cos^2(\eta{/2}) \, (\sin^2\theta)/4   \nonumber \\
\big| \big\langle  (0, 1) \, 1  \, 2 \, ; 1 \, 0 \big| \tilde{0} , \tilde{0},  \tilde{0} \big\rangle \big|^2
&=& \sin^2(\eta{/2}) \, (\sin^2\theta)/4   \times (2/5)   \nonumber \\
\big| \big\langle  (1, 0) \, 1  \, 2 \, ; 1 \, 0 \big| \tilde{0} , \tilde{0},  \tilde{0} \big\rangle \big|^2
&=& \sin^2(\eta{/2}) \, (\sin^2\theta)/4   \times (2/5)    ,
\end{eqnarray*}
for $J=2, M=0$
\begin{eqnarray*}
\big| \big\langle  (1, 1) \, 2  \, 0 \, ; 2 \, 0 \big| \tilde{0} , \tilde{0},  \tilde{0} \big\rangle \big|^2
&=& \cos^2(\eta{\/2}) \, \sin^4(\theta{/2})  \times (2/3) \nonumber \\
\big| \big\langle  (0, 0) \, 0  \, 2 \, ; 2 \, 0 \big| \tilde{0} , \tilde{0},  \tilde{0} \big\rangle \big|^2
&=& \sin^2(\eta{/2}) \, \cos^4(\theta{/2})   \nonumber \\
\big| \big\langle  (1, 1) \, 0  \, 2 \, ; 2 \, 0 \big| \tilde{0} , \tilde{0},  \tilde{0} \big\rangle \big|^2
&=& \sin^2(\eta{/2}) \, \sin^4(\theta{/2}) \times (1/3)   \nonumber \\
\big| \big\langle  (1, 1) \, 2  \, 2 \, ; 2 \, 0 \big| \tilde{0} , \tilde{0},  \tilde{0} \big\rangle \big|^2
&=& \sin^2(\eta{/2}) \, \sin^4(\theta{/2})  \times (4/21)    ,
\end{eqnarray*}
for $J=3, M=0$
\begin{eqnarray*}
\big| \big\langle  (0, 1) \, 1  \, 2 \, ; 3 \, 0 \big| \tilde{0} , \tilde{0},  \tilde{0} \big\rangle \big|^2
&=& \sin^2(\eta{/2}) \, (\sin^2\theta)/4 \times (3/5)  \nonumber \\
\big| \big\langle  (1, 0) \, 1  \, 2 \, ; 3 \, 0 \big| \tilde{0} , \tilde{0},  \tilde{0} \big\rangle \big|^2
&=& \sin^2(\eta{/2}) \, (\sin^2\theta)/4 \times (3/5)     ,
\end{eqnarray*}
and for $J=4, M=0$
\begin{eqnarray*}
\big| \big\langle  (1, 1) \, 2  \, 2 \, ; 4 \, 0 \big| \tilde{0} , \tilde{0},  \tilde{0} \big\rangle \big|^2
=\sin^2(\eta{/2}) \, \sin^4(\theta{/2})  \times (12/35)   .
\end{eqnarray*}
When ${\cal E}$ and $\tilde{r}$ are such that ${E}_2 < {E}_1$, one has to perform the switching
$\cos^2(\eta{/2}) \leftrightarrow \sin^2(\eta{/2})$
in the expressions above.
When ${\cal E}=0$, $\theta=\eta=0$ and only the term with both cosines
survives, namely the one in Eq.~\eqref{P0-B} for $J=0$.

\subsection{For indistinguishable fermions}
For fermions, when ${\cal E}$ and $\tilde{r}$ are such that ${E}_2 \ge {E}_1$, the probabilities are  \\

\noindent for $J=0, M=0$
\begin{eqnarray*}
\big| \big\langle  (0, 1) \, 1  \, 1 \, ; 0 \, 0 \big| \tilde{0} , \tilde{0},  \tilde{1} \big\rangle \big|^2
&=& \cos^2(\eta{/2})  \, (\sin^2\theta)/4 \times (1/3)   \nonumber \\
\big| \big\langle  (1, 0) \, 1  \, 1 \, ; 0 \, 0 \big| \tilde{0} , \tilde{0},  \tilde{1} \big\rangle \big|^2
&=& \cos^2(\eta{/2})  \, (\sin^2\theta)/4  \times (1/3) ,
\end{eqnarray*}
for $J=1, M=0$
\begin{eqnarray}
\big| \big\langle  (0, 0) \, 0  \, 1 \, ; 1 \, 0 \big| \tilde{0} , \tilde{0},  \tilde{1} \big\rangle \big|^2
&=& \cos^2(\eta{/2})  \, \cos^4(\theta{/2})     \\
\big| \big\langle  (1, 1) \, 0  \, 1 \, ; 1 \, 0 \big| \tilde{0} , \tilde{0},  \tilde{1} \big\rangle \big|^2
&=& \cos^2(\eta{/2})  \, \sin^4(\theta{/2}) \times (1/3)   \nonumber \\
\big| \big\langle  (1, 1) \, 2  \, 1 \, ; 1 \, 0 \big| \tilde{0} , \tilde{0},  \tilde{1} \big\rangle \big|^2
&=& \cos^2(\eta{/2})  \, \sin^4(\theta{/2})  \times (4/15)   \nonumber \\
\big| \big\langle  (1, 1) \, 2  \, 3 \, ; 1 \, 0 \big| \tilde{0} , \tilde{0},  \tilde{1} \big\rangle \big|^2
&=& \sin^2(\eta{/2})  \, \sin^4(\theta{/2}) \times { (6/35) } \nonumber  ,
\label{P1-F}
\end{eqnarray}
for $J=2, M=0$
\begin{eqnarray*}
\big| \big\langle  (0, 1) \, 1  \, 1 \, ; 2 \, 0 \big| \tilde{0} , \tilde{0},  \tilde{1} \big\rangle \big|^2
&=&  \cos^2(\eta{/2})  \, (\sin^2\theta)/4  \times (2/3)  \nonumber \\
\big| \big\langle  (1, 0) \, 1  \, 1 \, ; 2 \, 0 \big| \tilde{0} , \tilde{0},  \tilde{1} \big\rangle \big|^2
&=&  \cos^2(\eta{/2})  \, (\sin^2\theta)/4  \times (2/3)  \nonumber \\
\big| \big\langle  (0, 1) \, 1  \, 3 \, ; 2 \, 0 \big| \tilde{0} , \tilde{0},  \tilde{1} \big\rangle \big|^2
&=& \sin^2(\eta{/2})  \, (\sin^2\theta)/4  \times (3/7)    \nonumber \\
\big| \big\langle  (1, 0) \, 1  \, 3 \, ; 2 \, 0 \big| \tilde{0} , \tilde{0},  \tilde{1} \big\rangle \big|^2
&=&  \sin^2(\eta{/2})  \, (\sin^2\theta)/4  \times (3/7)     ,
\end{eqnarray*}
for $J=3, M=0$
\begin{eqnarray*}
\big| \big\langle  (1, 1) \, 2  \, 1 \, ; 3 \, 0 \big| \tilde{0} , \tilde{0},  \tilde{1} \big\rangle \big|^2
&=& \cos^2(\eta{/2}) \, \sin^4(\theta{/2})  \times (2/5) \nonumber \\
\big| \big\langle  (0, 0) \, 0  \, 3 \, ; 3 \, 0 \big| \tilde{0} , \tilde{0},  \tilde{1} \big\rangle \big|^2
&=& \sin^2(\eta{/2}) \, \cos^4(\theta{/2})   \nonumber \\
\big| \big\langle  (1, 1) \, 0  \, 3 \, ; 3 \, 0 \big| \tilde{0} , \tilde{0},  \tilde{1} \big\rangle \big|^2
&=&  \sin^2(\eta{/2}) \, \sin^4(\theta{/2})  \times (1/3)  \nonumber \\
\big| \big\langle  (1, 1) \, 2  \, 3 \, ; 3 \, 0 \big| \tilde{0} , \tilde{0},  \tilde{1} \big\rangle \big|^2
&=& \sin^2(\eta{/2}) \, \sin^4(\theta{/2}) \times (8/45)     ,
\end{eqnarray*}
for $J=4, M=0$
\begin{eqnarray*}
\big| \big\langle  (0, 1) \, 1  \, 3 \, ; 4 \, 0 \big| \tilde{0} , \tilde{0},  \tilde{1} \big\rangle \big|^2
= \sin^2(\eta{/2})  \, (\sin^2\theta)/4  \times (4/7)  \nonumber \\
\big| \big\langle  (1, 0) \, 1  \, 3 \, ; 4 \, 0 \big| \tilde{0} , \tilde{0},  \tilde{1} \big\rangle \big|^2
= \sin^2(\eta{/2})  \, (\sin^2\theta)/4  \times (4/7)   ,
\end{eqnarray*}
and for $J=5, M=0$
\begin{eqnarray*}
\big| \big\langle  (1, 1) \, 2  \, 3 \, ; 5 \, 0 \big| \tilde{0} , \tilde{0},  \tilde{1} \big\rangle \big|^2
= \sin^2(\eta{/2}) \, \sin^4(\theta{/2})  \times (20/63)  .
\end{eqnarray*}
When ${\cal E}$ and $\tilde{r}$ are such that ${E}_2 < {E}_1$, one has to perform the switching
$\cos^2(\eta{/2}) \leftrightarrow \sin^2(\eta{/2})$
in the expressions above.
When ${\cal E}=0$, $\theta=\eta=0$ and only the term with both cosines
survives, namely the one in Eq.~\eqref{P1-F} for $J=1$.

\section{QDT parameters for collisions of fermionic NaK}\label{app:d}

\begin{figure}[h]
\begin{center}
\includegraphics*[width=12cm, trim=0cm 4cm 1cm 4cm]{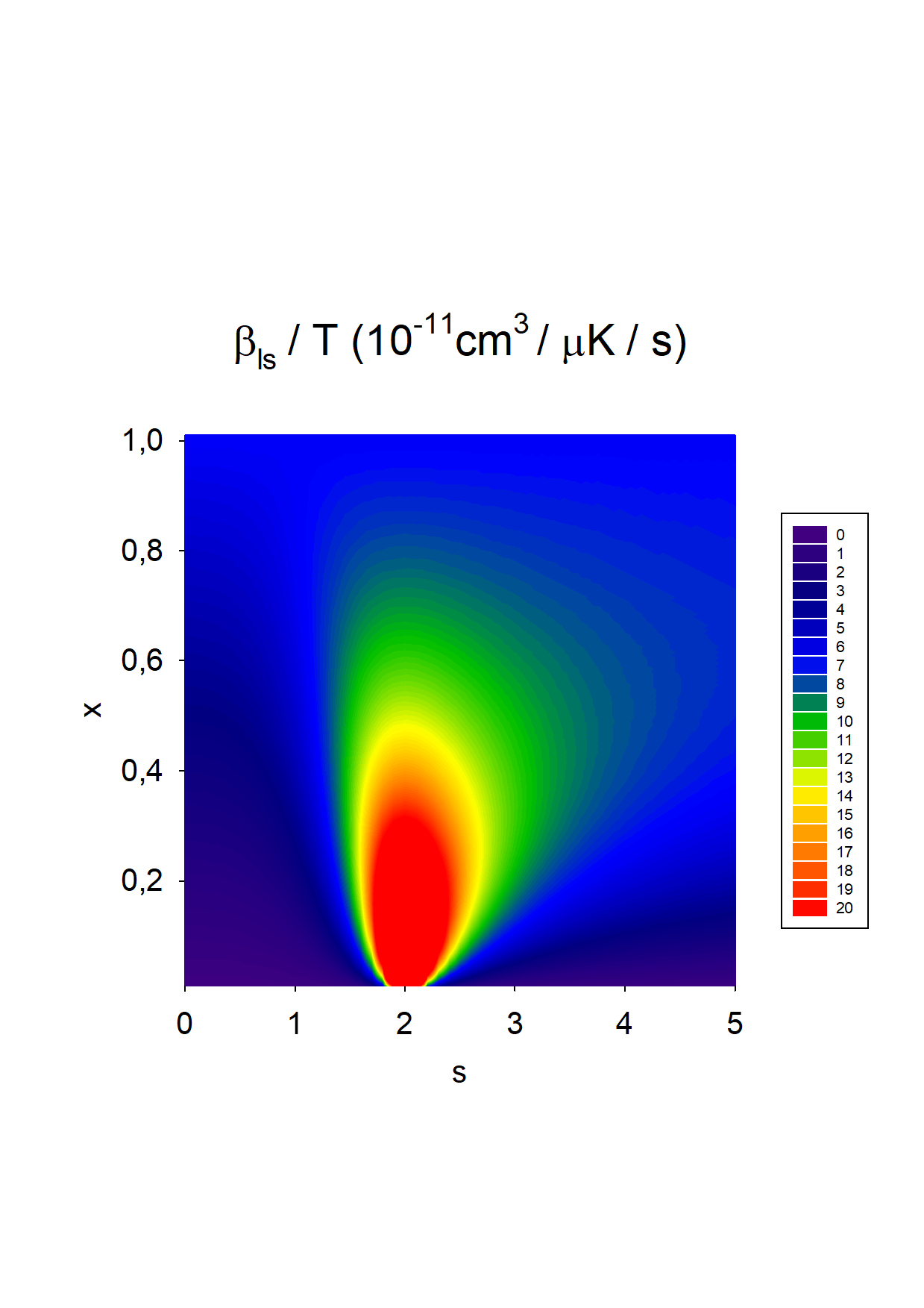}
\caption{Value of the molecular loss slope
$\beta_\text{ls}/T$ in $10^{-11}$ cm$^3$/$\mu$K/s
as a function of the two QDT parameter $s$ and $x$ 
present in the theoretical formalism 
developped in \cite{Idziaszek_PRL_104_113202_2010}.
The experimental value found in \cite{Bause_PRR_3_033013_2021}
corresponds to $\beta_\text{ls}/T \sim 13. \,10^{-11}$ cm$^3$/$\mu$K/s,
represented in light green in the picture (see color code).}
\label{FIG-QDT-NAK}
\end{center}
\end{figure}

We used the QDT theoretical formalism of \cite{Idziaszek_PRL_104_113202_2010}
to plot the molecular loss slope $\beta_\text{ls}/T$
as a function of the two QDT parameters $s$ and $x$.
This is presented in Fig.~\ref{FIG-QDT-NAK}.
Note that the present $x$ parameter replaces the $y$ parameter in \cite{Idziaszek_PRL_104_113202_2010} 
to be consistent with the notation of the 
unified model in \cite{Croft_PRA_102_033306_2020}.
To fit the experimental molecular 
loss slope of $\beta_\text{ls}/T \sim 13. \,10^{-11}$ cm$^3$/$\mu$K/s found in \cite{Bause_PRR_3_033013_2021}
at zero electric field,
we found that we need values of $x \le 0.5$ with a range $1.6 \le s \le 2.7$.

\twocolumngrid

\end{document}